\documentclass[aps,prd,showpacs,floatfix,preprintnumbers,nofootinbib,amssymb,amsmath]{revtex4}
\usepackage{graphicx,bm,color}
\usepackage{subfigure}

\newcommand{\idp}{\int_\lambda^\Lambda \frac{d^4p}{(2\pi)^4} }
\newcommand{\Tr}{\mathrm{Tr}}
\newcommand{\cC}{{\cal C}}
\newcommand{\cT}{{\cal T}}
\newcommand{\cM}{{\cal M}}

\begin{document}
\title{Thermodynamic properties of strongly interacting matter in a finite volume 
using the Polyakov-Nambu-Jona-Lasinio model }
\author{Abhijit Bhattacharyya}
\email{abphy@caluniv.ac.in}
\affiliation{Department of Physics, University of Calcutta,
92, A. P. C. Road, Kolkata - 700009, INDIA}
\author{Paramita Deb}
\email{paramita.deb83@gmail.com}
\affiliation{Theory Division, Physical Research Laboratory, Navarangpura, Ahmedabad - 380009,
 INDIA}
\author{Sanjay K. Ghosh}
\email{sanjay@bosemain.boseinst.ac.in}
\affiliation{Department of Physics and Centre for Astroparticle Physics \&
Space Science, Bose Institute,
93/1, A. P. C Road, Kolkata - 700009, INDIA}
\author{Rajarshi Ray}
\email{rajarshi@bosemain.boseinst.ac.in}
\affiliation{Department of Physics and Centre for Astroparticle Physics \&
Space Science, Bose Institute,
93/1, A. P. C Road, Kolkata - 700009, INDIA}
\author{Subrata Sur}
\email{ssur.phys@gmail.com}
\affiliation{Department of Physics, Panihati Mahavidyalaya, Barasat Road, Sodepur, 
Kolkata - 700110, INDIA}

\vskip 0.3in
\begin {abstract}
 We present the thermodynamic properties of strongly interacting matter
in finite volume in the framework of Polyakov loop enhanced 
Nambu$-$Jona-lasinio model within mean field approximation. We considered
both the 2 flavor and  2+1 flavor matter. Our primary observation was
a qualitative change in the phase transition properties that resulted in
the lowering of the temperature corresponding to the critical end point.
This would make it favorable for detection in heavy-ion experiments that
intend to create high density matter with considerably small
temperatures. We further demonstrate the possibility of obtaining 
chiral symmetry restoration even within the confined phase in finite
volumes.

\end{abstract}
\pacs{12.38.Aw, 12.38.Mh, 12.39.-x}

\maketitle

\section {Introduction}

The strongly interacting matter is supposed to have a rich phase
structure at finite temperatures and densities~\cite{rajagopal}.
While our Universe at the present epoch contains a significant fraction
of color singlet hadrons, color non-singlet states especially quarks
and gluons may have been prevalent in the past $-$ a few microseconds
after the Big Bang~\cite{kolb}. The temperature of the Universe at
that epoch is estimated to be $\sim 200$ MeV. Similar state of matter
is also expected to exist inside the core of super-massive stars in the
present day Universe, where the density is $\sim 10$ times that of
normal nuclear matter. A direct study of such natural phenomenon
is out of bounds even to modern astrophysicists. Fortunately experimental
facilities at CERN (France/Switzerland), BNL (USA) and recently at
GSI (Germany) are exploring the possibilities of creating and
studying the properties of such exotic states of matter in a controlled
environment. The key differences that appear in such experiments as
compared to the natural phenomenon are the lifetime of matter created
in the exotic state and its volume. Whereas in natural phenomenon the
lifetime of the exotic matter may be large compared to the interaction
time-scale, it is usually very small in an experimental situation.
Some effects of the enhanced lifetime on the physical aspects of the
system relating to the onset of equilibrium of weak interactions were
discussed by us in Ref.~\cite{beta}. Here we shall discuss about the effects
of finite volume on the properties of strongly interacting matter.

 In the following we shall generically define the matter with color
confined states as the hadronic phase and the exotic state with colored
degrees of freedom as the quark-gluon plasma (QGP). In the experiments
this exotic phase may be produced by ultra relativistic
collisions of heavy ions. The volume of the system thus created would
depend on the nature of the colliding nuclei, the center of mass energy
($\sqrt s$) and the centrality of collision. Once created, the system
expands until the constituents are so far separated that their 
interaction ceases and they flow out as free streaming particles.
The distribution of particles thus freezes out, except for some further
decays to smaller particles.
There have been a large number of efforts to estimate the system size
at freeze-out for different $\sqrt s$ and different centralities. The
most popular way of doing so is to measure the Hanbury-Brown-Twiss radii 
(see e.g. \cite{boalke,lisapratt} for reviews). In Ref.~\cite{ceres} 
it has been shown that the freeze out volume increases as
the $\sqrt s$ increases. Here the authors have estimated the
freeze out volume and found it to vary from $2000~fm^3$ to
$3000~fm^3$. In a very recent paper~\cite{graf} the volume of
homogeneity has been calculated using UrQMD model~\cite{urqmd} and
compared with the
experimentally available results. The $\sqrt s$ considered was in the
range of 62.4 GeV to 2760 GeV for lead-lead collisions at different
centralities. The system volume has been found to vary from $50~fm^3$ to  
$250~fm^3$. Given that these are the freeze-out volumes, one can 
trace back to the initial equilibration time and expect an even smaller
system size. In fact one cannot even consider the whole fireball, which
is an isolated system to be in thermodynamic equilibrium. One has to
choose a proper rapidity interval to act as the system under
consideration. Therefore it becomes important to study how the various
thermodynamic quantities in a strongly interacting matter depend on the
volume of the system. Specifically we know that finite system sizes
would lead to smoothening of any singularities appearing at a phase
transition~\cite{fsize}. Thus important signatures of such transitions
must be reanalyzed with the help of finite size scaling analysis~\cite{fss}. 
In the context of heavy ion collisions such a possible
analysis has been discussed in the literature
(see e.g.~\cite{fss1,fss2,fss3}).

 On the theoretical side a study of finite volume effects was done in
Ref.~\cite{elze} with a bag of non-interacting quarks and gluons and
it was found that the effective degrees of freedom are reduced. In
Ref.~\cite{spieles} a two model equation of state was used to show that
the separation between the hadronic and QGP phases around the
critical temperature looses its sharpness resulting in a soft effective
equation of state. A few first principle study of pure gluon theory on
space-time lattices were performed, showing the possibility of
significant finite size effects~\cite{fslat1,fslat2}. Similar studies
are going on in various QCD inspired models. In
Ref.~\cite{fischer1,fischer2} the quark mass gap equation has been
studied with Schwinger-Dyson equation parallel to equivalent Lattice
QCD (LQCD) calculations and various meson properties are found to have
significant volume dependence. In the context of chiral perturbation
theory the implications of finite system size have been discussed
\cite{lusher1,gasser1}. Then there are studies with four-fermi type
interactions, like the Nambu$-$Jona-Lasinio (NJL) \cite{Nambu} models
~\cite{kiriyama,fss1,shao}, linear
sigma models~\cite{fss2,braun1,braun2} and Gross-Neveu models~\cite{khanna}. 
While in Ref.~\cite{braun1} the scaling behavior of chiral phase
transition for finite and infinite volumes has been studied, the character
of phase diagram has been studied in Ref.~\cite{fss1,fss2,braun2,khanna}.
In refs.~\cite{kiriyama} and~\cite{shao} the authors have studied the chiral
properties as a function of the radius of a finite droplet of quark
matter. The stability of such a droplet in the context of strangelet
formation within the NJL model has been addressed in Ref.~\cite{yasui}.
Size dependent effects of difermion states within 2-dimensional NJL
model has been studied in Ref.~\cite{abreu1} and that of magnetic field is
discussed in Ref.~\cite{abreu2}. Recently in a 1+1 dimensional NJL model
the induction of charged pion condensation phenomenon in dense baryonic
matter due to finite volume effects have been studied in \cite{ebert}.

In this work we shall use the Polyakov loop enhanced NJL (PNJL) model
to study the thermodynamic properties of the strongly interacting matter
in a finite volume. This model originated from the NJL
model~\cite{klev,hat1,hat2} 
which incorporates the global symmetries of
QCD quite nicely. A four quark interaction term in the NJL Lagrangian is 
able to generate the physics of spontaneous breaking of chiral symmetry 
$-$ a property of QCD which is manifested as the non-degenerate chiral 
partners of the low-mass hadrons. However a reasonable description of
the physics of color confinement is missing. With the introduction of
a background field in the NJL model, motivated by the dynamics of the
Polyakov Loop~\cite{poly}, one obtains the PNJL model which describes
a number of features of confinement physics quite satisfactorily
(see e.g.~\cite{fuku,ratti1,megias1,ray1,ratti2,ray2,gatto,deb,megias2}).

Certain aspects of finite volume effects in the PNJL model has been
discussed in Ref.~\cite{cristo} through a coarse graining of the Lagrangian,
followed by a Monte Carlo simulation. This method goes on similar lines
as the numerical studies of LQCD. Normally this would involve the same
kind of complex determinant problem that has plagued the direct LQCD
computations for non-zero baryon number densities. So it may
be desirable to keep using the saddle point approximation in PNJL model
to study the finite volume effects. Here we make the first case study,
albeit with some simplified assumptions towards that direction.

  We organize our paper as follows. In the next section we briefly 
describe the PNJL model and the modifications for finite volume. In
section \ref{secphase} we describe phase
transition at finite volume and in section \ref{secthermo} we discuss
the thermodynamic properties. The pion and sigma meson masses and the
pion decay constant at finite volume have been discussed in section
\ref{meson}. In section \ref{secconclu} we summarize and conclude.

\section {The PNJL model}
\label{secmodel}

We shall consider the PNJL model with light flavors (2 flavor) and
light plus strange flavors (2+1 flavor). In the PNJL model
the gluon physics comes into play through the chiral point couplings
between quarks (present in the NJL part) and a background field which
represents Polyakov Loop dynamics. The Polyakov line is represented as,

\begin {equation}
  L(\bar x)={\cal P} {\rm exp}[i {\int_0}^\beta
d\tau A_4{({\bar x},\tau)}]
\end {equation}
where $A_4=iA_0$ is the temporal component of Eucledian gauge field
$(\bar A,A_4)$, $\beta=\frac {1}{T} $, and $\cal P$ denotes path
ordering. $L(\bar x)$ transforms as a field with charge one under
global Z(3) symmetry. The Polyakov loop is then given by 
$\Phi = (Tr_c L)/N_c$, and its conjugate by,
${\bar \Phi} = (Tr_c L^\dagger)/N_c$. The gluon dynamics can be
described as an effective theory of the Polyakov loops.
The Polyakov loop potential can be expressed as,
\begin{equation}
\frac {{\cal {U^\prime}}(\Phi[A],\bar \Phi[A],T)} {T^4}= 
\frac  {{\cal U}(\Phi[A],\bar \Phi[A],T)}{ {T^4}}-
                                     \kappa \ln(J[\Phi,{\bar \Phi}])
\label {uprime}
\end{equation}
where $\cal {U(\phi)}$ is a Landau-Ginsburg type potential commensurate
with the Z(3) global symmetry. Here we choose a form given
in Ref.~\cite{ratti1},
\begin{equation}
\frac  {{\cal U}(\Phi, \bar \Phi, T)}{  {T^4}}=-\frac {{b_2}(T)}{ 2}
                 {\bar \Phi}\Phi-\frac {b_3}{ 6}(\Phi^3 + \bar \Phi^3)
                 +\frac {b_4}{  4}{(\bar\Phi \Phi)}^2,
\end{equation}
where
\begin {eqnarray}
     {b_2}(T)=a_0+{a_1}(\frac { {T_0}}{ T})+{a_2}(\frac {{T_0}}{ T})^2+
              {a_3}(\frac {{T_0}}{T})^3,
\end {eqnarray}
$b_3$ and $b_4$ being constants. The second term in Eqn.(\ref{uprime})
is the Vandermonde term which replicates the effect of SU(3) Haar
measure and is given by,
\begin {equation}
J[\Phi, {\bar \Phi}]=(27/24{\pi^2})\left[1-6\Phi {\bar \Phi}+\nonumber\\
4(\Phi^3+{\bar \Phi}^3)-3{(\Phi {\bar \Phi})}^2\right]
\end{equation}
The corresponding parameters were earlier obtained in the above
mentioned literature by choosing suitable values by fitting a few
physical quantities as function of temperature obtained in LQCD
computations. The set of values chosen here are,
\vskip 0.1in
\begin{center}
$a_0=6.75$, $a_1=-1.95$, $a_2=2.625$, $a_3=-7.44$, $b_3=0.75$,
  $b_4=7.5$,  \\ 
$T_0=190 \, {\rm MeV}$, $\kappa=0.2$ (for 2 flavor), $\kappa = 0.13$
(for 2+1 flavor)
     \end{center}
\vskip 0.1in

 For the quarks we shall use the usual form of the NJL model except
for the substitution of a covariant derivative containing a background
temporal gauge field. Thus the 2 flavor version of PNJL model is
described by the Lagrangian,  
\begin {align}
   {\cal L} &= {\sum_{f=u,d}}{\bar\psi_f}\gamma_\mu iD^\mu
             {\psi_f}-\sum_f m_{f}{\bar\psi_f}{\psi_f}
              +\sum_f \mu_f \gamma_0{\bar \psi_f}{\psi_f}\nonumber\\
     &+{\frac {g_S} {2}} {\sum_{a=1,2,3}}[({\bar\psi} \tau^a {\psi})^2+
            ({\bar\psi} i\gamma_5\tau^a {\psi})^2] 
        -{\cal {U^\prime}}(\Phi[A],\bar \Phi[A],T)
\label{lag1}
\end {align}
For 2+1 flavor the Lagrangian may be written as,
\begin {align}
   {\cal L} = {\sum_{f=u,d,s}}{\bar\psi_f}\gamma_\mu iD^\mu
             {\psi_f}&-\sum_f m_{f}{\bar\psi_f}{\psi_f}
              +\sum_f \mu_f \gamma_0{\bar \psi_f}{\psi_f}
       +{\frac {g_S} {2}} {\sum_{a=0,\ldots,8}}[({\bar\psi} \lambda^a
        {\psi})^2+
            ({\bar\psi} i\gamma_5\lambda^a {\psi})^2] \nonumber\\
       &-{g_D} [det{\bar\psi_f}{P_L}{\psi_{f^\prime}}+det{\bar\psi_f}
            {P_R}{\psi_{f^\prime}}]
                 -{\cal {U^\prime}}(\Phi[A],\bar \Phi[A],T)
\label{lag2}
\end {align}
where $f$ denotes the flavors $u$ or $d$ or $s$ respectively.
The matrices $P_{L,R}=(1\pm \gamma_5)/2$ are respectively the
left-handed and right-handed chiral projectors, and the other terms
have their usual meaning, described in details in
refs.~\cite{ray1,ray2,deb,deb2,deb3,deb4}. This NJL part of the theory
is analogous to the BCS theory of superconductor, where the
pairing of two electrons leads to the condensation causing a gap in
the energy spectrum. Similarly in the chiral limit, NJL model exhibits
dynamical breaking of ${SU(N_f)}_L \times {SU(N_f)_R}$ symmetry to
$SU(N_f)_V$ symmetry ($N_f$ being the number of flavors). As a result 
the composite operators ${\bar \psi_f}\psi_f$ pick up nonzero vacuum
expectation values. The quark condensate is given as,
\begin {equation}
 \langle{\bar \psi_f}{\psi_f}\rangle= 
-i{N_c}{{{\cal L}t}_{y\rightarrow x^+}}(tr {S_f}(x-y)),
\end {equation}
where trace is over color and spin states. The self-consistent gap 
equation for the constituent quark masses are,
\begin {equation}
  M_f =m_f-g_S \sigma_f+g_D \sigma_{f+1}\sigma_{f+2},
\end {equation}
where $\sigma_f=\langle{\bar \psi_f} \psi_f\rangle$ denotes chiral 
condensate of the quark with flavor $f$. Here if we consider
$\sigma_f=\sigma_u$, then $\sigma_{f+1}=\sigma_d$ and
$\sigma_{f+2}=\sigma_s$. Similarly if $\sigma_f=\sigma_d$ then
$\sigma_{f+1}=\sigma_s$ and $\sigma_{f+2}=\sigma_u$,
if $\sigma_f=\sigma_s$ then $\sigma_{f+1}=\sigma_u$ and 
$\sigma_{f+2}=\sigma_d$. The expression for $\sigma_f$ at zero
temperature ($T=0$) and chemical potential ($\mu_f=0$) may be
written as~\cite{gatto},
\begin {equation}
 \sigma_f=-\frac {3{M_f}}{ {\pi}^2} {{\int}^\Lambda}\frac {p^2}{
           \sqrt {p^2+{M_f}^2}}dp,
\end {equation}
$\Lambda$ being the three-momentum cut-off. This cut-off have been used
to regulate the model because it contains dimensionful couplings
 rendering the model to be non-renormalizable.

Due to the dynamical breaking of chiral symmetry, $N_f^2 - 1$ Goldstone
bosons appear. These are the pions and kaons whose masses, decay widths
etc. from experimental observations are utilized to fix the NJL model
parameters. The parameter values have been listed in table \ref{table1}.
Here we consider the $\Phi$, $\bar \Phi$ and $\sigma_f$ fields in the
mean field approximation (MFA) where the mean field are obtained by
simultaneously solving the respective saddle point equations.

\begin{table}[htb]
\begin{center}
\begin{tabular}{|c|c|c|c|c|c|}
\hline
Model&$ m_u $&$ m_s $&$ \Lambda $&$ g_S \Lambda^2 $&$ g_D \Lambda^5 $\\
& MeV & MeV & MeV &  & \\
\hline
2 flavor& 5.5 &0 &651 & 4.27 &0 \\
2+1 flavor& 5.5 & 134.76 & 631 & 3.67 & 9.33\\
\hline
\end{tabular}
\caption{Parameters of the Fermionic part of the model.}  
\label{table1}
\end{center}
\end{table}

 Now that the PNJL model is described for infinite volumes we discuss
how we implement the finite volume constraints. Ideally one should
choose the proper boundary conditions $-$ periodic for bosons and
anti-periodic for fermions. This would lead to a infinite sum over
discrete momentum values $p_i=\pi n_i/R$, where $i=x,y,z$ and $n_i$
are all positive integers and $R$ is the lateral size of a cubic
volume. This implies a lower momentum cut-off
$p_{min}=\pi/R=\lambda$ (say). One should also incorporate proper
effects of surface and curvatures. In this first case study we shall
however take up a number of simplifications listed below:

\noindent
(i) We shall neglect surface and curvature effects.

\noindent
(ii) The infinite sum will be considered as an integration over a continuous
variation of momentum albeit with the lower cut-off.

\noindent
(iii) We shall not use any modifications to the mean-field parameters due to
finite size effects. Our philosophy had been to hold the known physics
at zero $T$, zero $\mu$ and infinite $V$ fixed. That means we treat $V$
as a thermodynamic variable in the same footing as $T$ and $\mu$.
Therefore any variation due to change in either of these thermodynamic
parameters were translated into the changes in the effective fields of
$\sigma_f$, $\Phi$ etc. and through them to the meson spectra. The values
of meson masses and decay constants used to fix the model parameters
were thus naturally expected to be the values strictly at $T=0$ and
$\mu=0$ and $V=\infty$. Thus the Polyakov loop potential as well as the
mean-field part of the NJL model would remain unchanged. They shall 
feel the effect of changing volume only implicitly through the saddle
point equations.

\section {Phase Transition}
\label{secphase}

To study the finite volume effects on the thermodynamic properties of
strongly interacting matter we begin by writing down the thermodynamic
potential in MFA.

\begin{figure}[htb]
\centering
\includegraphics[scale=0.5]{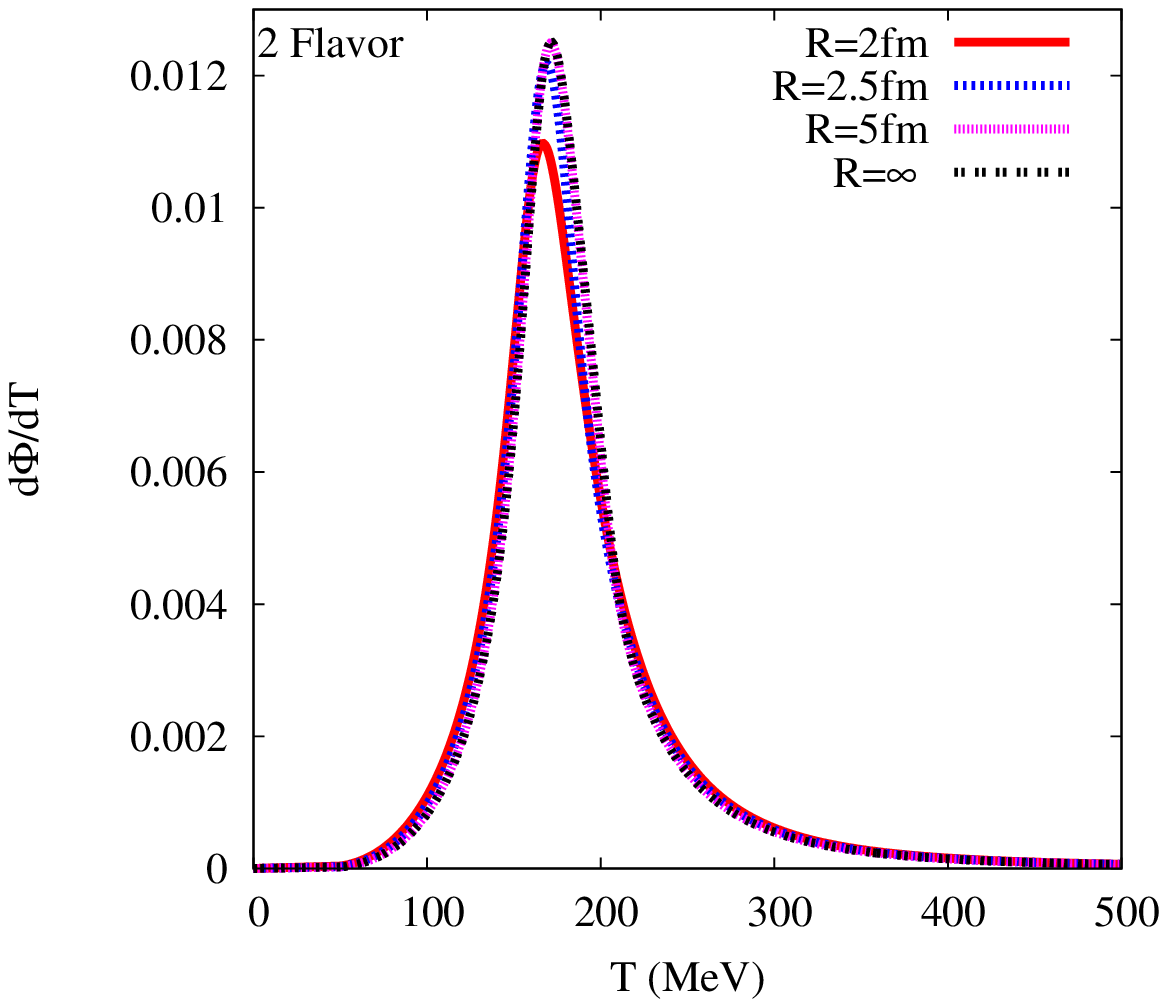}
\includegraphics[scale=0.5]{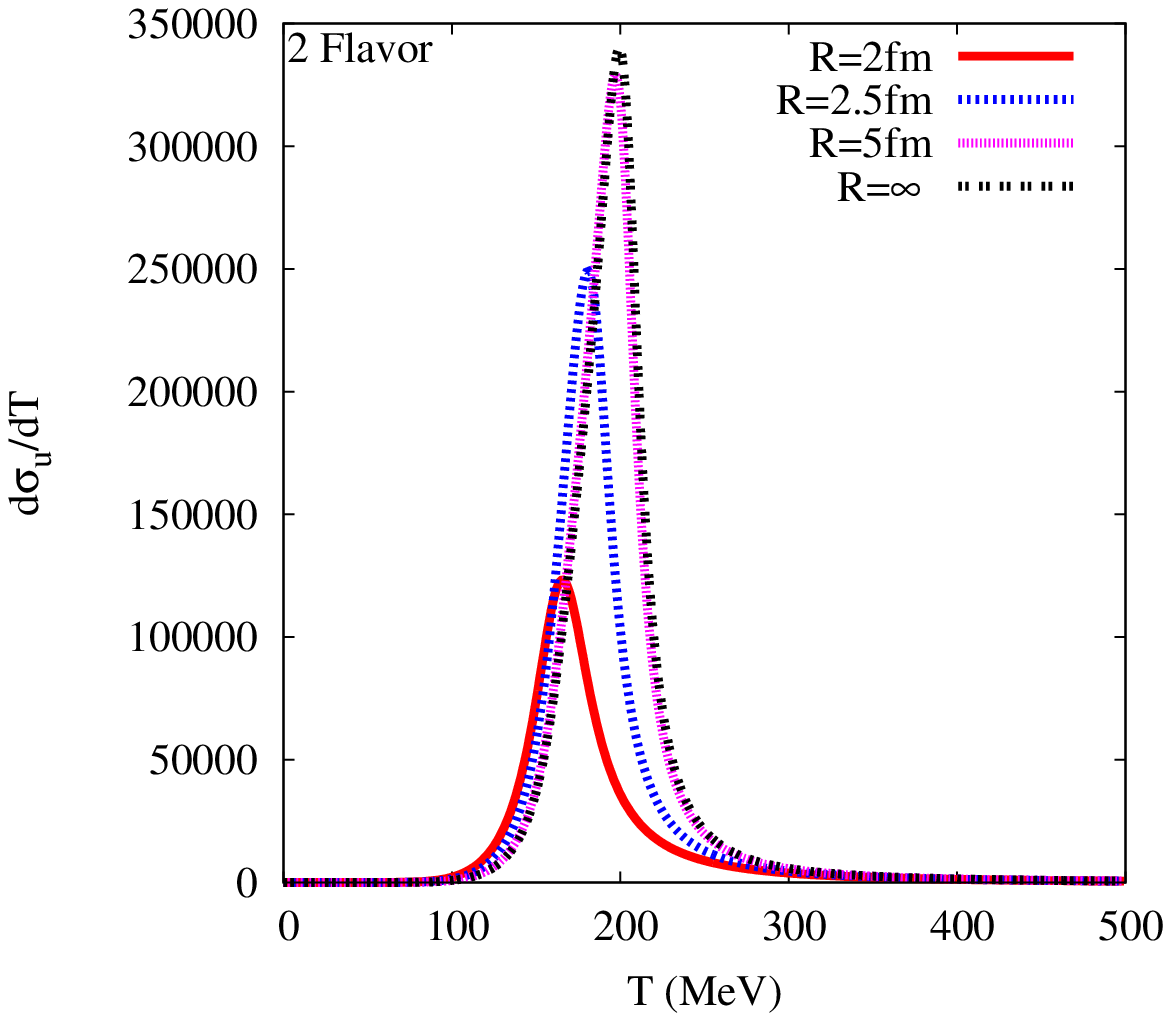}
\includegraphics[scale=0.5]{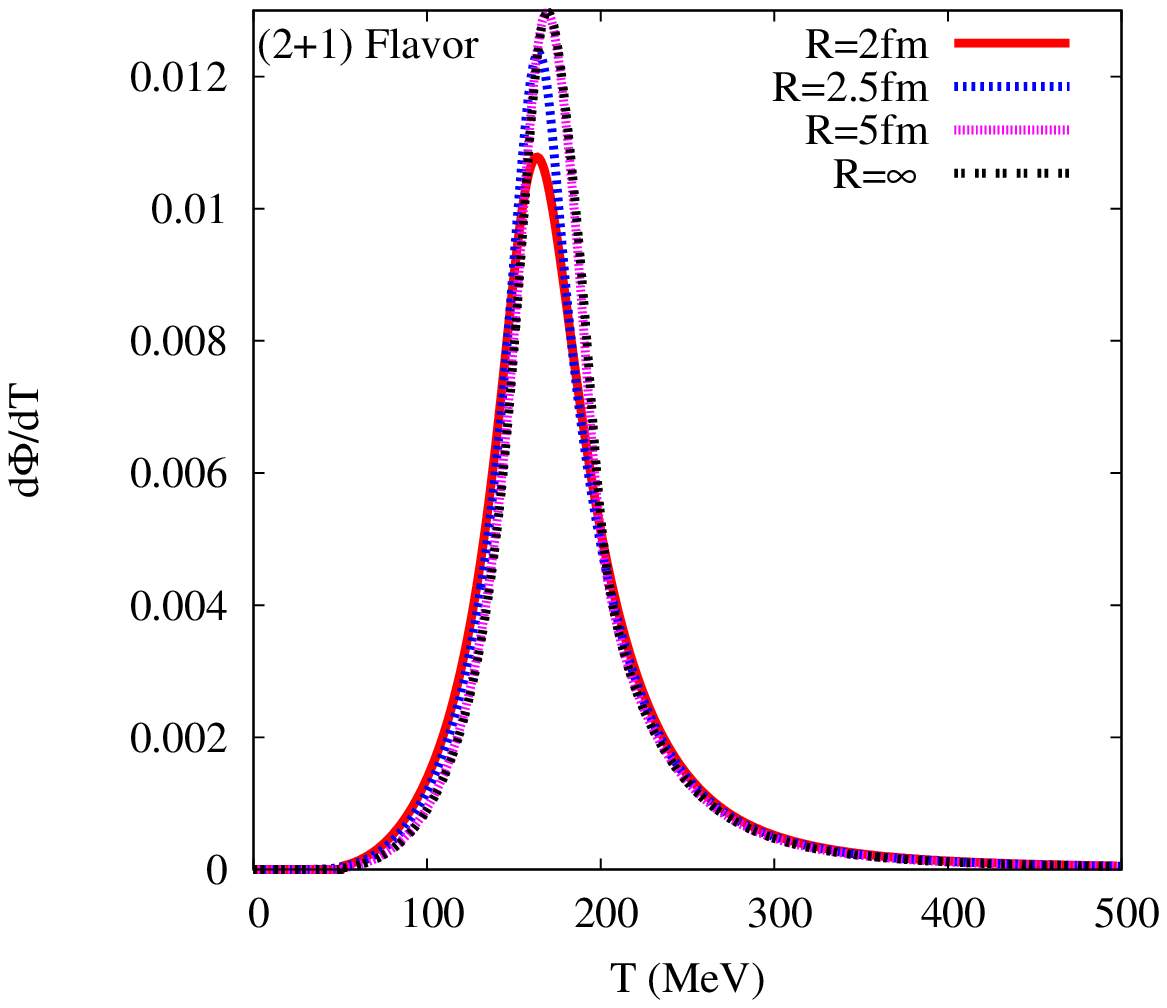}
\includegraphics[scale=0.5]{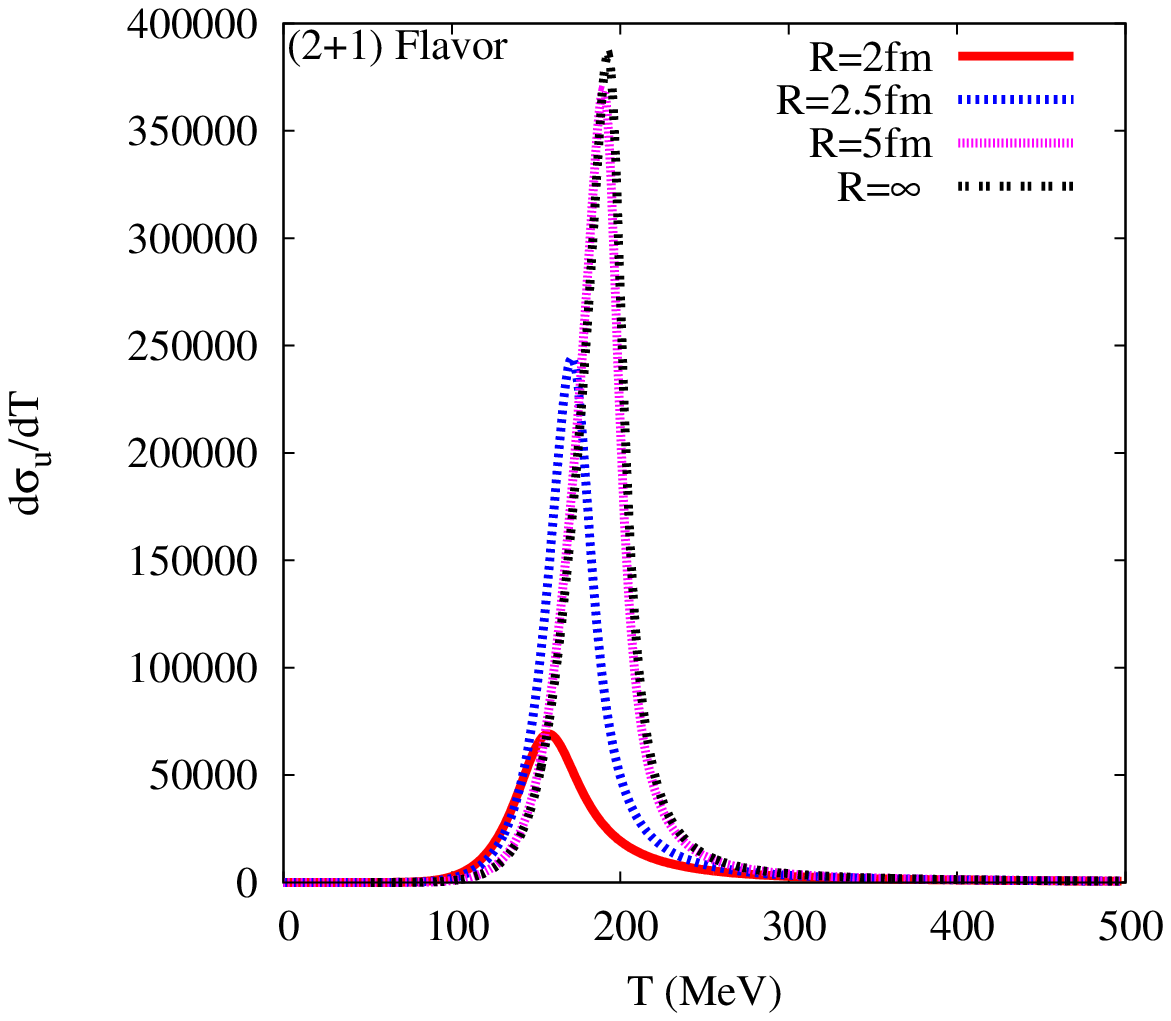}
\caption{(Color online) Derivatives of order parameters for chiral and 
deconfinement phase transition for different system sizes.}
\label {deriv}
\end{figure}

The expression is given by,
\begin {align}
  \Omega^\prime (\Phi,\bar \Phi,\sigma_f,T,\mu_f)&= {\cal {U^\prime}}
       [\Phi,\bar \Phi,T] + 2{g_S}{\sum_{f=u,d,s}}
            {\sigma_f^2}-\frac {{g_D}}{ 2}{\sigma_u}
          {\sigma_d}{\sigma_s} \nonumber \\
  &-T{\sum_n}\int_{\lambda}^{\infty}\frac {{d^3p}}{{(2{\pi})^3}}
                {Tr} \ln\frac {{ S^{-1}}(i{\omega_n},\bar p)}{ T}.
\end {align}
where $\omega_n=\pi T(2n+1)$ are Matsubara frequencies for fermions.
The inverse quark propagator is given in momentum space by
\begin {equation}
 {S^{-1}} =
     \gamma_0(p^0+{\hat \mu}-i{A_4})-\vec{\gamma}\cdot{\vec p}-{\hat M}
\end {equation}
using the identity ${Tr}\ln\left(X\right)=\ln \det\left(X\right)$, we
get,
\begin {align}
 \Omega^\prime &= {\cal {U^\prime}}[\Phi,\bar \Phi,T]
    + 2{g_S} {\sum_{f=u,d,s}} {\sigma_f^2}
    - \frac {{g_D} }{2}{\sigma_u}{\sigma_d}{\sigma_s}
    - 6 {\sum_f}{\int_{\lambda}^{\Lambda}}
     \frac {{d^3p}}{{(2\pi)}^3} E_{p_f}\Theta {(\Lambda-{ |\vec p|})}
                                                           \nonumber \\
       &-2{\sum_f}T{\int_\lambda^\infty}\frac {{d^3p}}{{(2\pi)}^3}
     \ln\left[1+3(\Phi+{\bar \Phi}\exp(\frac{-(E_{p_f}-\mu_f)}{T}))
   \exp(\frac{-(E_{p_f}-\mu_f)}{T})+\exp(\frac{-3(E_{p_f}-\mu_f)}{T})
        \right]
                                                            \nonumber\\
       &-2{\sum_f}T{\int_\lambda^\infty}\frac {{d^3p}}{{(2\pi)}^3}
     \ln\left[1+3({\bar \Phi}+{\Phi}\exp(\frac{-(E_{p_f}+\mu_f)}{T}))
   \exp(\frac{-(E_{p_f}+\mu_f)}{T})+\exp(\frac{-3(E_{p_f}+\mu_f)}{T})
        \right]
                                                           \nonumber \\
       &=\Omega - \kappa T^4 \ln J[\Phi,{\bar \Phi}]
\end {align}
where $E_{p_f}=\sqrt {p^2+M^2_f}$ is the single quasiparticle energy.
In the last line $\Omega$ contains all the terms of $\Omega^{\prime}$
except the Vandermonde term.

We now search for the saddle point of the thermodynamic potential which
gives the temperature and density dependence of the fields. For all the
system sizes, at zero baryon density, we found that the order
parameters for both chiral
$(\sigma=<{\bar u} u> + <{\bar d} d>)$ and deconfinement  $(\Phi)$
transition smoothly passes from the hadronic phase to the quark phase. 
This indicates that the system does not have a real phase transition,
rather there is a smooth crossover. The crossover temperature is
identified to be the point of inflection of $\sigma_u$ and $\Phi$ with
temperature. In Fig. \ref{deriv} we have plotted $d\Phi/dT$ and
$d\sigma_u/dT$ for 2 flavor and 2+1 flavor matter for different system
sizes. The peak position of these plots give respective inflection
points. Note that, the deconfinement and chiral transitions do not take
place exactly at the same temperature. Here we take the average of these
two temperatures as $T_c$. The results are shown in  table \ref{table2},
where we quote the different values of the crossover temperatures
corresponding to different system sizes.   

\begin{table}[htb]
\begin{center}
\begin{tabular}{|c|c|c|c|c|c|}
\hline
&$ R=2~fm $&$ R=2.5~fm $&$ R=3~fm $&$ R=5~fm $&$ R=\infty $\\
\hline
$T_c$ (MeV) (2 flavor)& 167& 171 & 180 &184& 186\\
$T_c$ (MeV) (2+1 flavor)&160 &167 &174&180 &181\\
\hline
\end{tabular}
\caption{Transition temperatures for different system sizes. }
\label{table2}
\end{center}
\end{table}

From table \ref{table2} it can be seen that the $T_c$ has a strong
dependence on system size. For 2 flavors the $T_c$ varies from $167$ MeV
to $186$ MeV which means a change of about $10\%$. A similar result is
observed for 2+1 flavor. One should note that the shift in
the $T_c$ is mainly due to the shift in the transition temperature of
the chiral phase transition. The transition temperature of the
deconfining phase transition almost does not change. This result is 
similar to that obtained with PNJL model on the lattice~\cite{cristo}.
This is somewhat expected as the Polyakov loop potential feels the
effect of changing volume only indirectly through the fields $\Phi$
and ${\bar \Phi}$.

\begin{figure}[htb]
\centering
\includegraphics[scale=0.5]{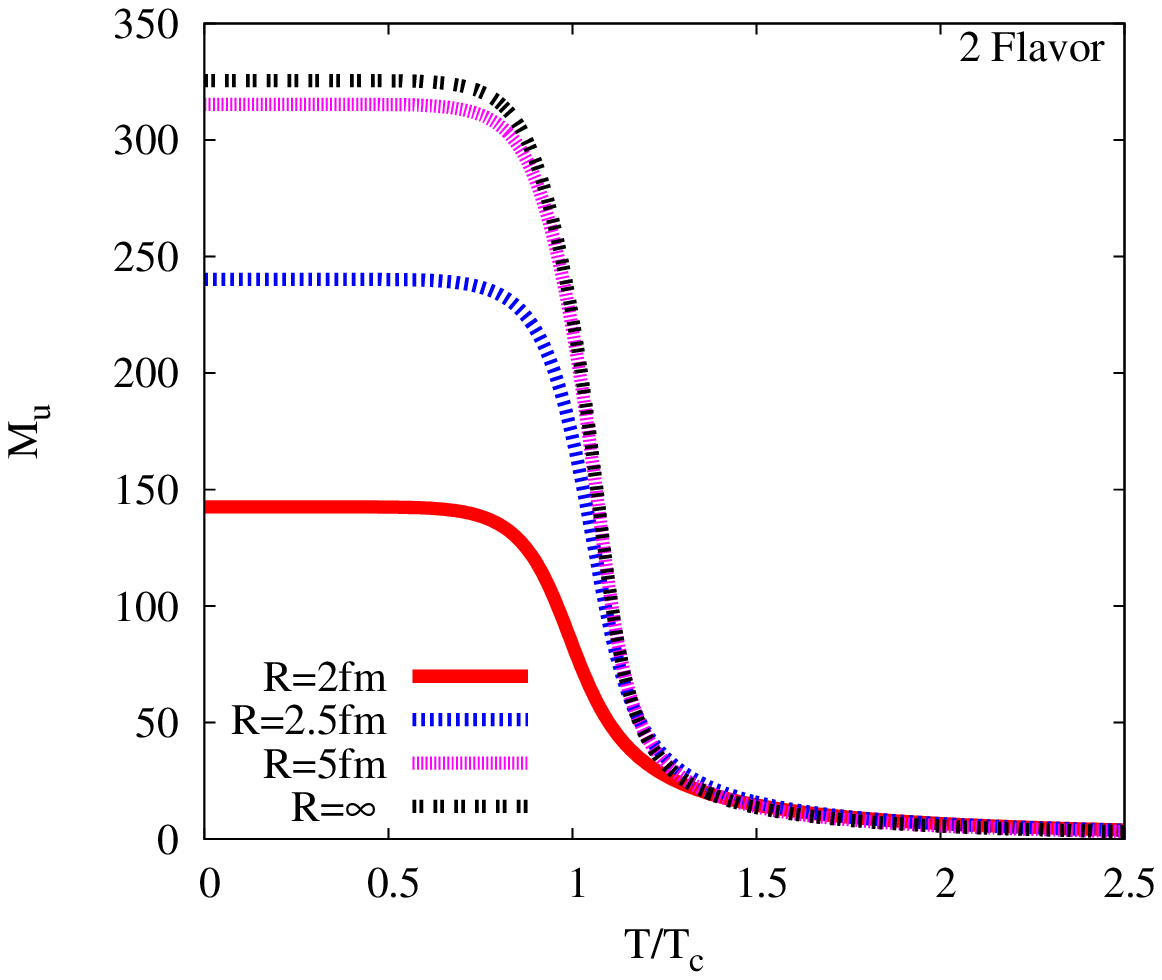} \\
\includegraphics[scale=0.5]{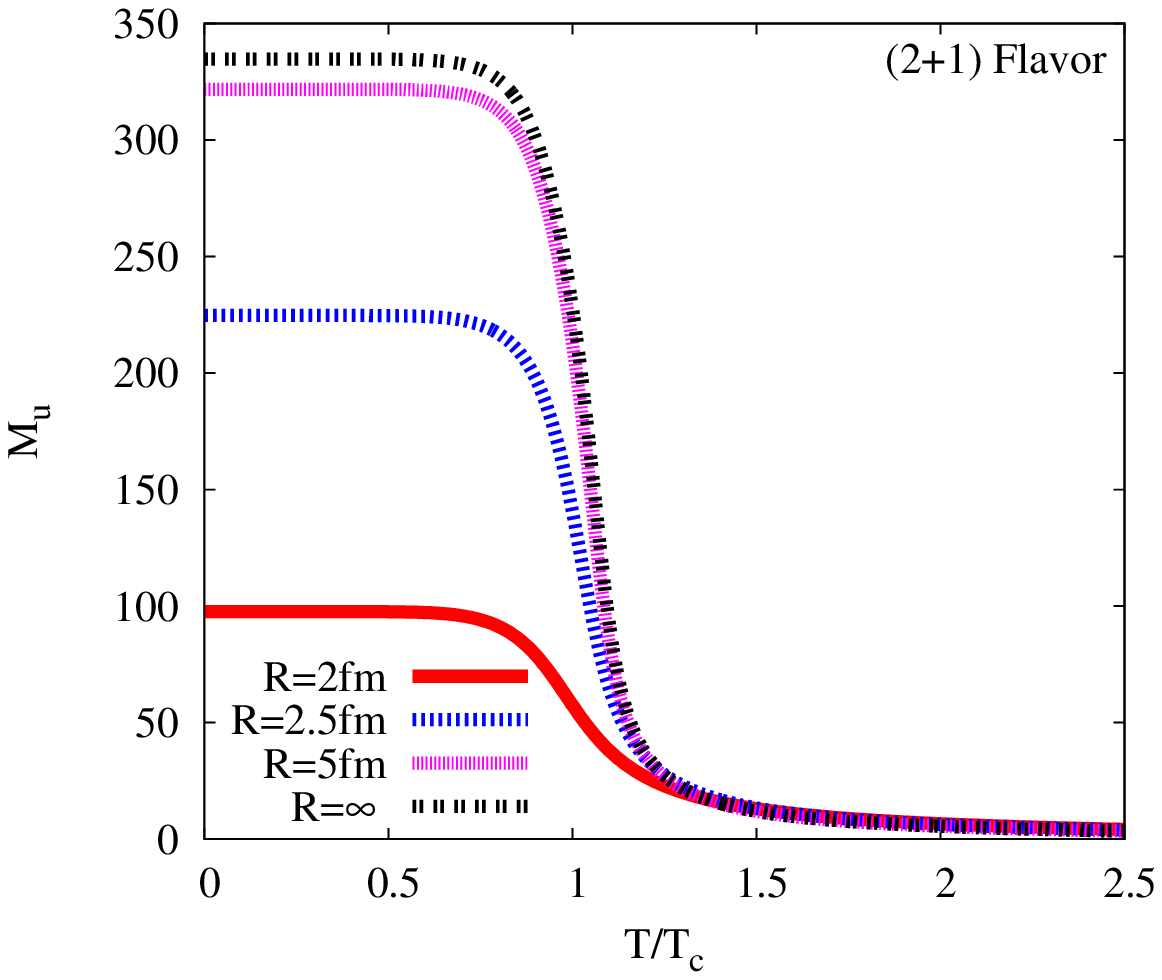}
\includegraphics[scale=0.5]{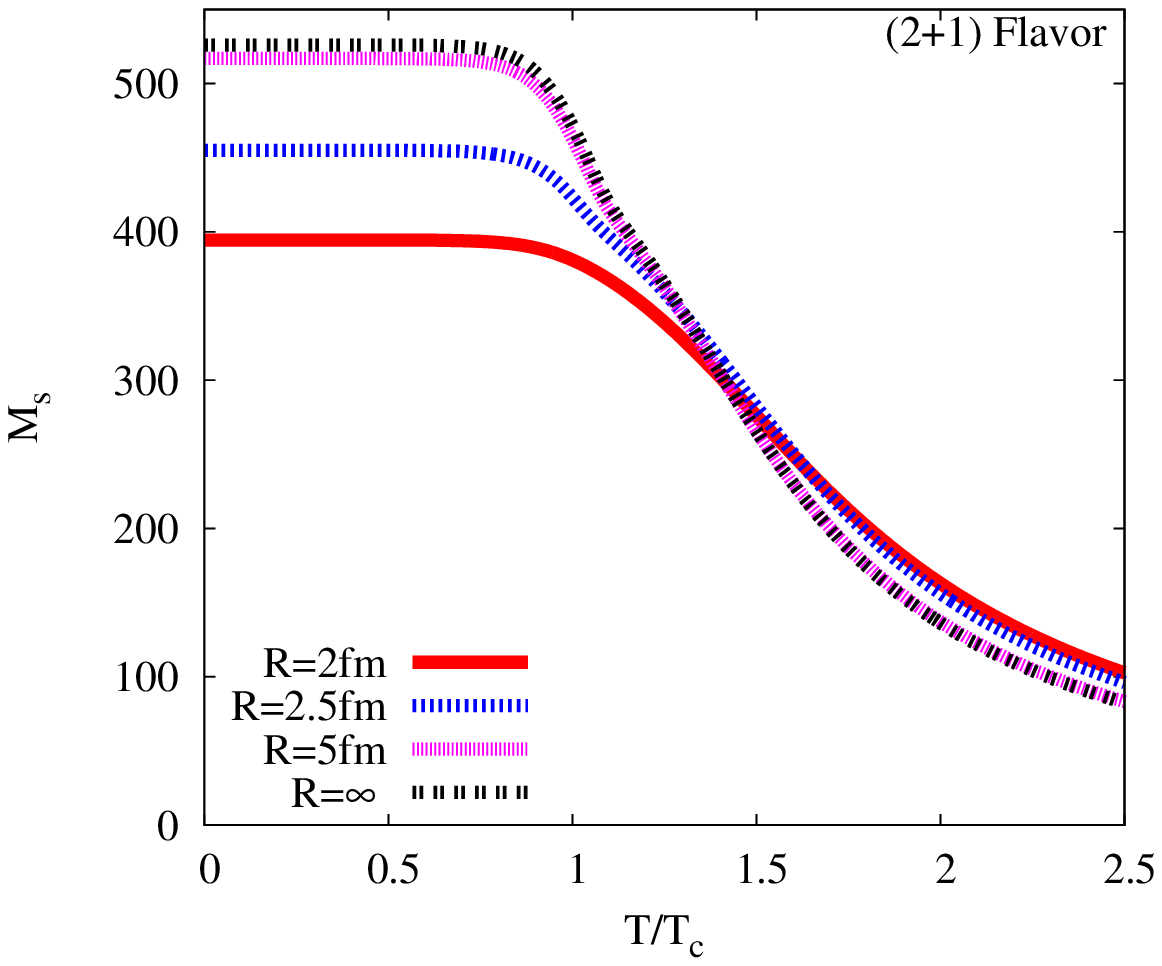}
\caption{(Color online) Constituent masses of quarks as a function of
temperature for different system sizes.}
\label{fig_mass}
\end{figure}

In Fig. \ref{fig_mass} we have plotted the temperature dependence of
the constituent quark masses for both 2 flavors and 2+1 flavors.
Below the crossover temperature they exhibit very strong volume
dependence. Smaller the volume, smaller is the constituent mass.
In the 2+1 flavor case, the masses of the light flavors drop faster
than the strange quark.
It thus seems that the chiral symmetry is gradually getting restored
as one looks into smaller and smaller volumes. This is also the
reason why the $T_c$ itself is lowered for smaller volumes given
in table \ref{table2}. Similar feature has also been observed
in NJL models~\cite{kiriyama,shao}. Given that the quark condensation
is similar to the superconducting condensate it is interesting to note
that there are in fact certain superconductors which show similar
decrease of band gap with the system size~\cite{supcond}.

\begin{figure}[htb]
\centering
\includegraphics[scale=0.5]{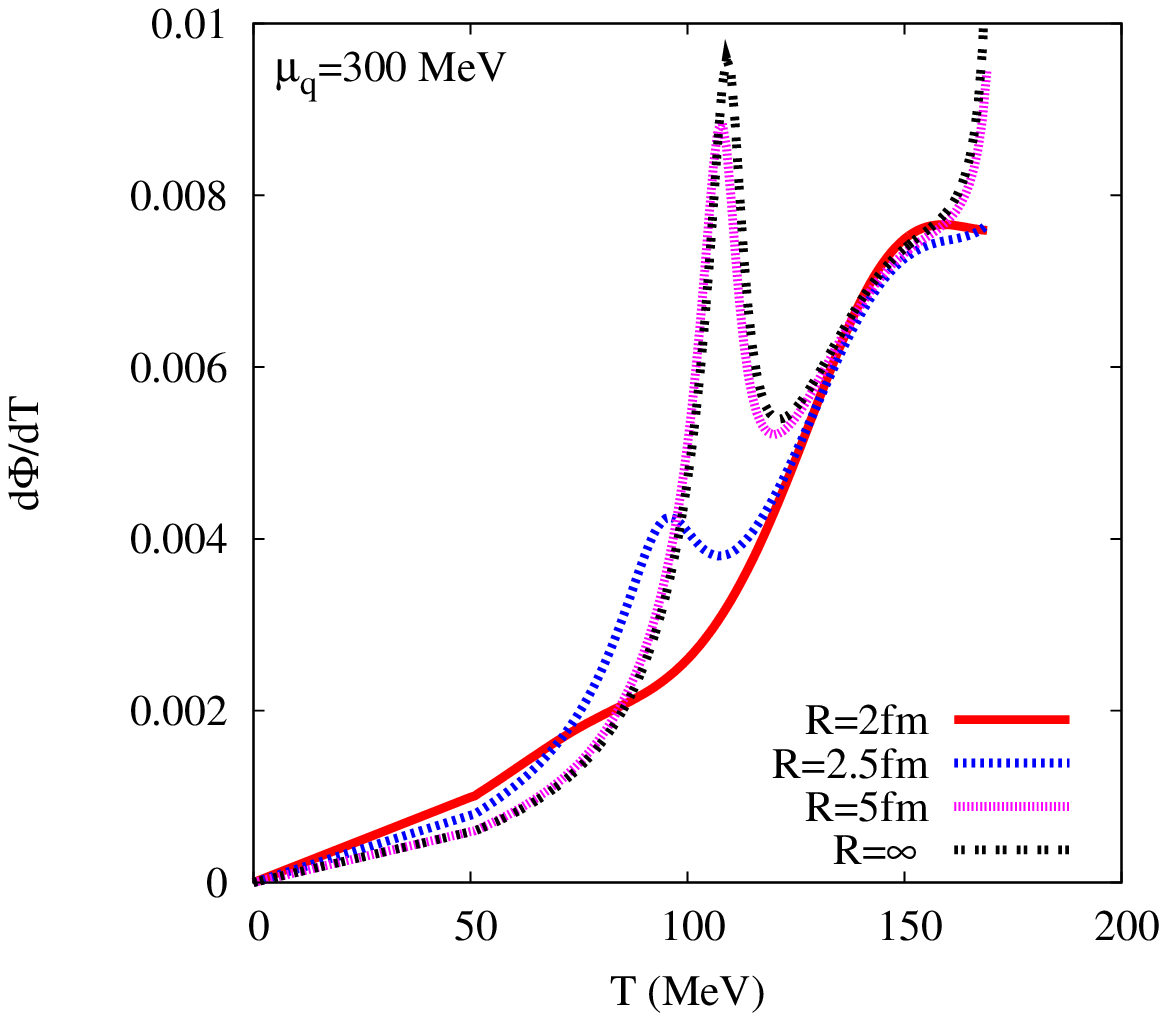}
\includegraphics[scale=0.5]{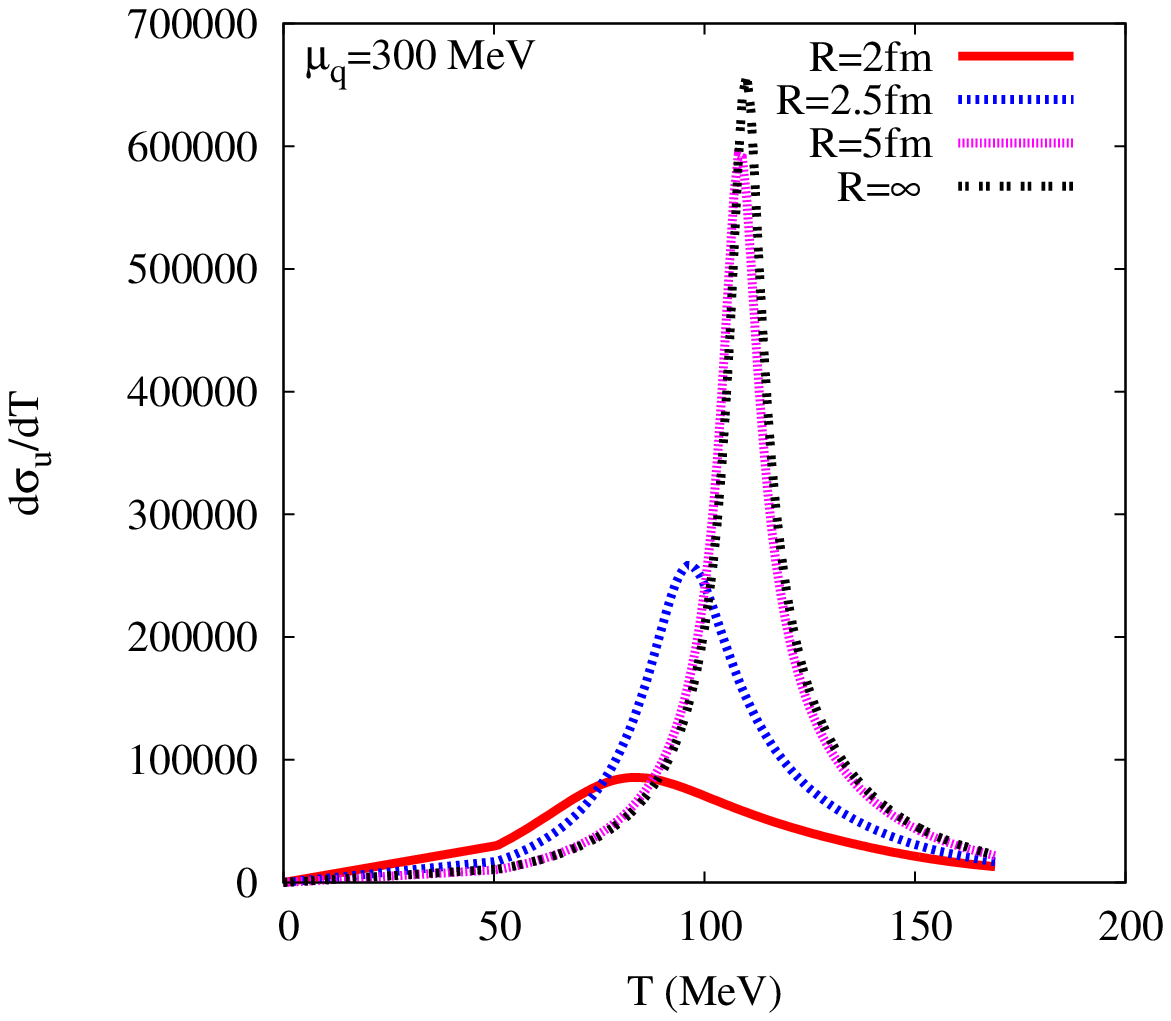}
\caption{(Color online) Derivatives of order parameters for chiral and 
deconfinement phase transition for 2 flavor at finite $\mu_q$.}
\label {derivmu}
\end{figure}

\begin{figure}[htb]
\centering
\includegraphics[scale=0.5]{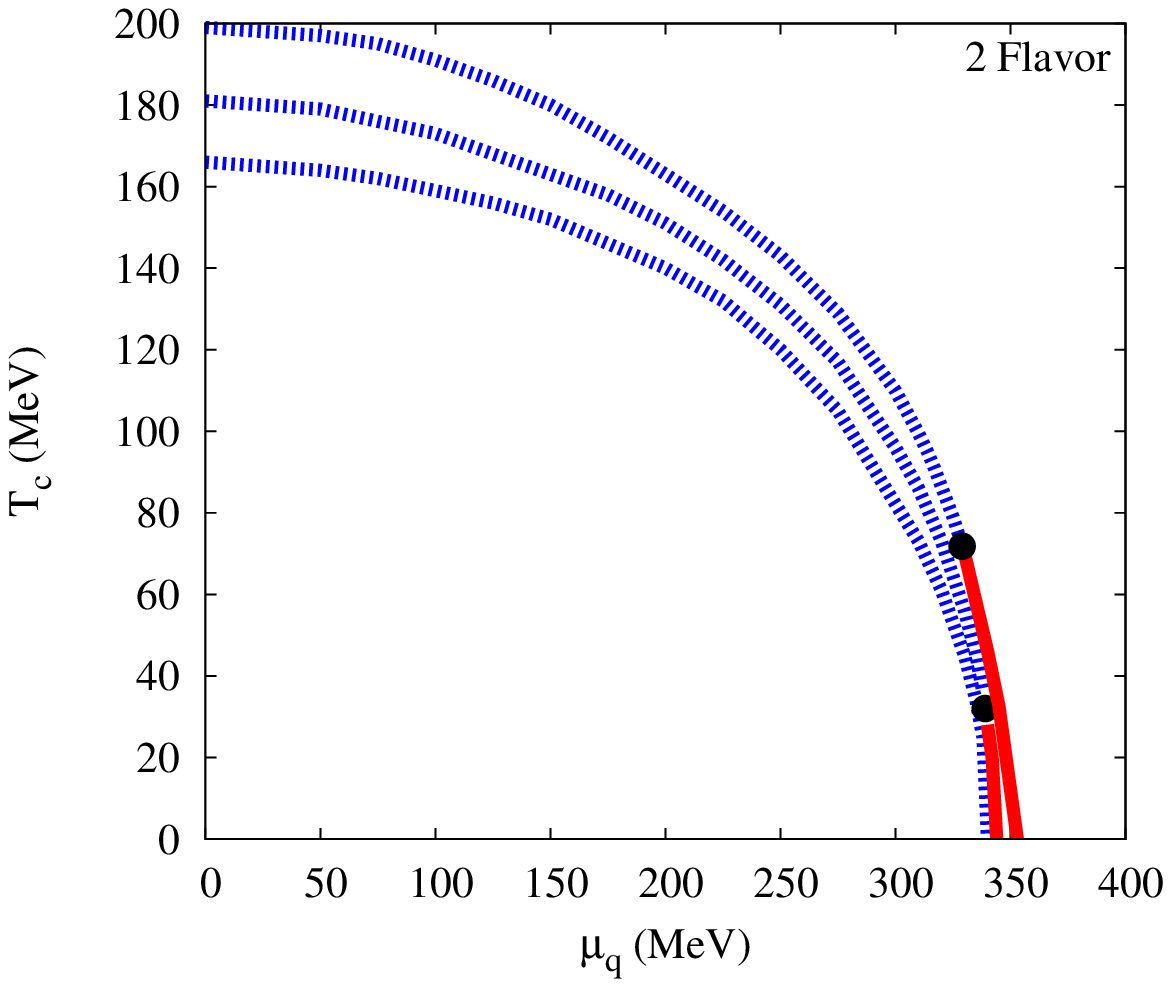}
\includegraphics[scale=0.5]{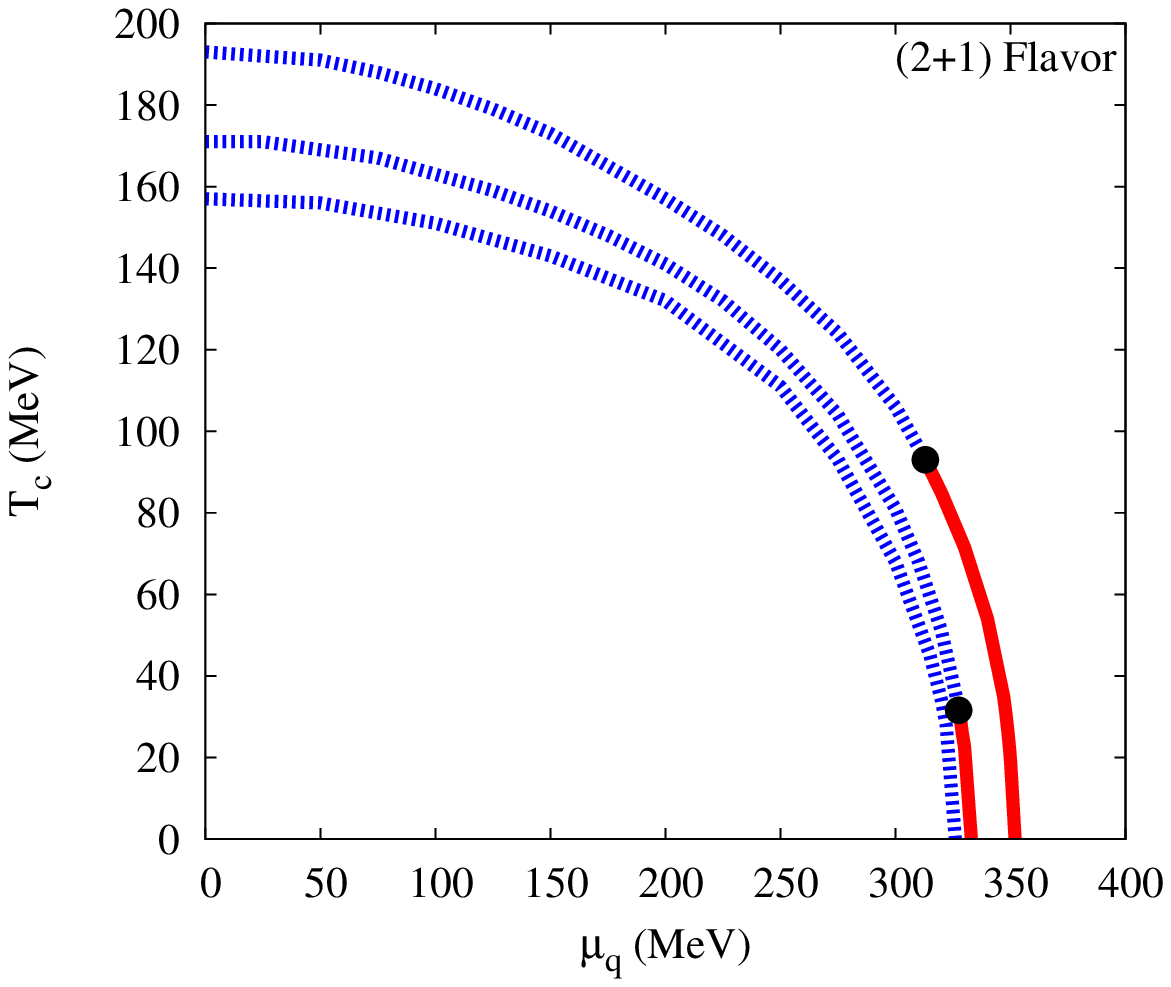}
\caption{(Color online) Phase diagram for different system sizes. The
inside curve is for $R=2~fm$, the next curve is for $R=2.5~fm$ and the
outermost curve is for $R=\infty$.}
\label {cep}
\end {figure}

Let us now take a look into the situation at non-zero quark chemical
potential $\mu_q=\sum_f \mu_f/N_f$. For infinite volume the phase
transition is of first
order and one observes a gap in the order parameter at sufficiently high
chemical potential. At some smaller $\mu_q$, the first order transition
ends at a critical end point (CEP). At this point the system undergoes a
second order transition. At even smaller $\mu_q$ we have only a
crossover. As the volume of the system is lowered we find the phase
transition characteristics fade away. Even the crossover characteristics
start to die down. This is clear from the Fig. \ref{derivmu} where we
plot $d\sigma_u/dT$ and $d\Phi/dT$ for $\mu_q = 300$ MeV as a function of
temperature. In Fig. \ref{cep} the phase diagram as a function of
system size is shown. Note that the CEP gradually shifts towards higher
$\mu_q$ and lower $T$ and finally disappears as the volume is reduced.
This is an encouraging fact for the critical point search in heavy-ion
collision experiments. To attain such high densities one needs to
collide the ions at low $\sqrt s$, which means the temperature attained
is lower. So if it were an infinite system one would have been far
away from the CEP. Fortunately the experiments would produce small
system volumes and this may lead to the location of the respective CEP
possible. Thereafter one would need to do the finite size scaling
analysis to extrapolate to the CEP for infinite volumes. The location
of CEP for different volumes is collected in table \ref{table3}.

\begin{table}[htb]
\begin{center}
\begin{tabular}{|c|c|c|c|c|c|}
\hline
&$ R=2~fm $&$ R=2.5~fm $&$ R=3~fm $&$ R=5~fm $&$ R=\infty $\\
\hline
$T_c$ (MeV), ${\mu_q}_c$ (MeV) (2 flavor)& No CEP & 32, 339 & 52, 335
&69, 330&72, 329\\
$T_c$ (MeV), ${\mu_q}_c$ (MeV) (2+1 flavor)& No CEP & 32, 328& 60, 324
&86, 316&93, 313\\
\hline
\end{tabular}
\caption{Location of chiral CEP for different system sizes. }
\label{table3}
\end{center}
\end{table}

\section{Thermodynamics}
\label{secthermo}

In this section we discuss the behavior of a few thermodynamic
observables namely pressure, energy density, specific heat, speed of
sound etc. for different system sizes.

\begin{figure}[htb]
\centering
\includegraphics[scale=0.5]{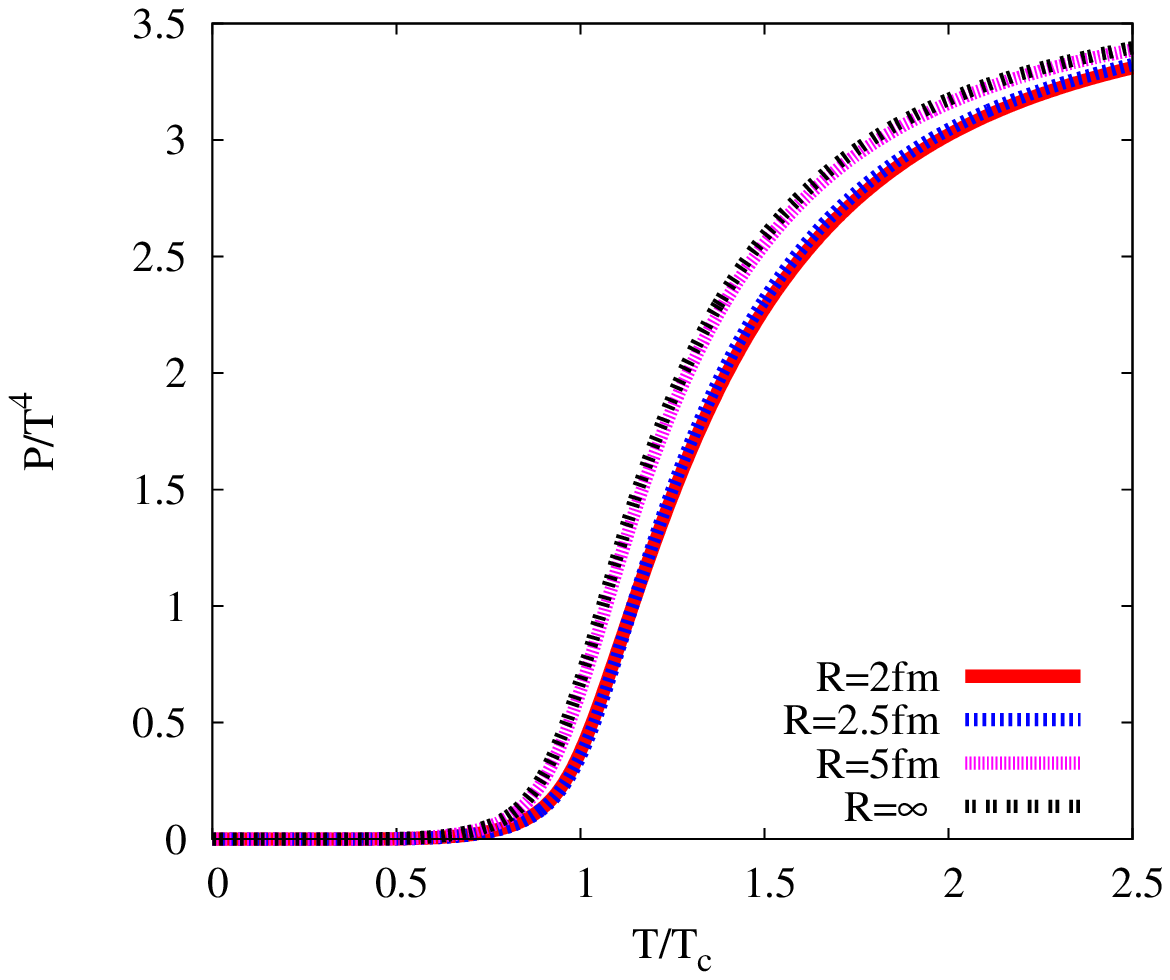}
\includegraphics[scale=0.5]{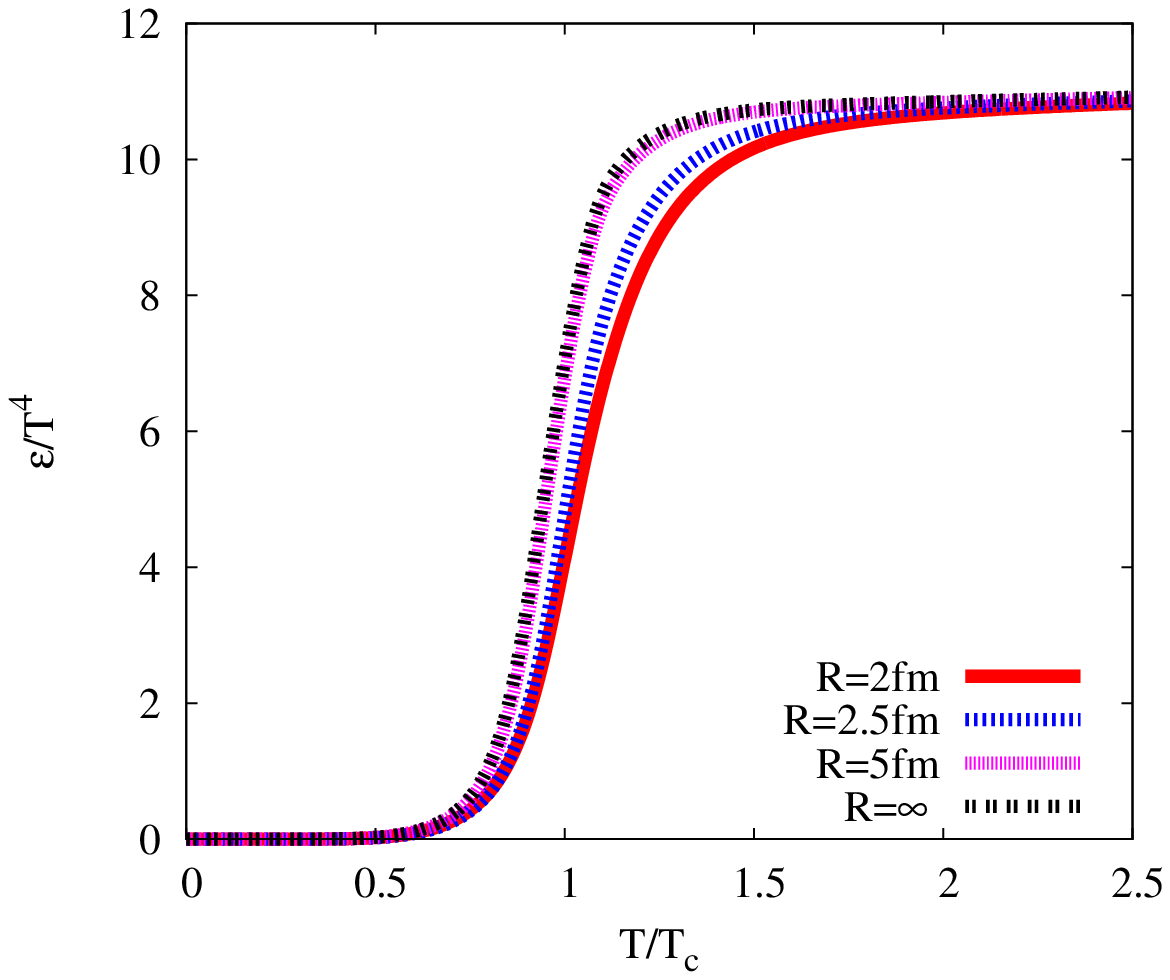}
\includegraphics[scale=0.5]{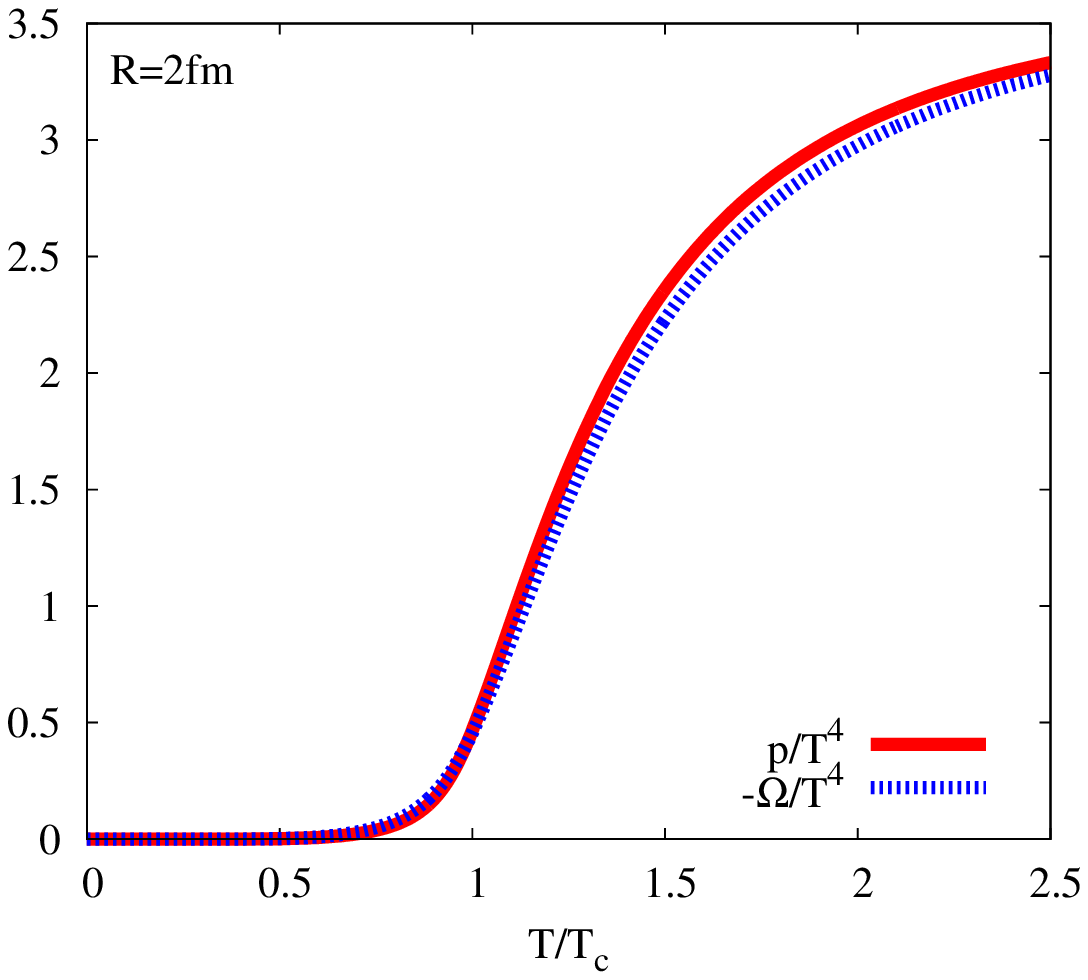}
\includegraphics[scale=0.5]{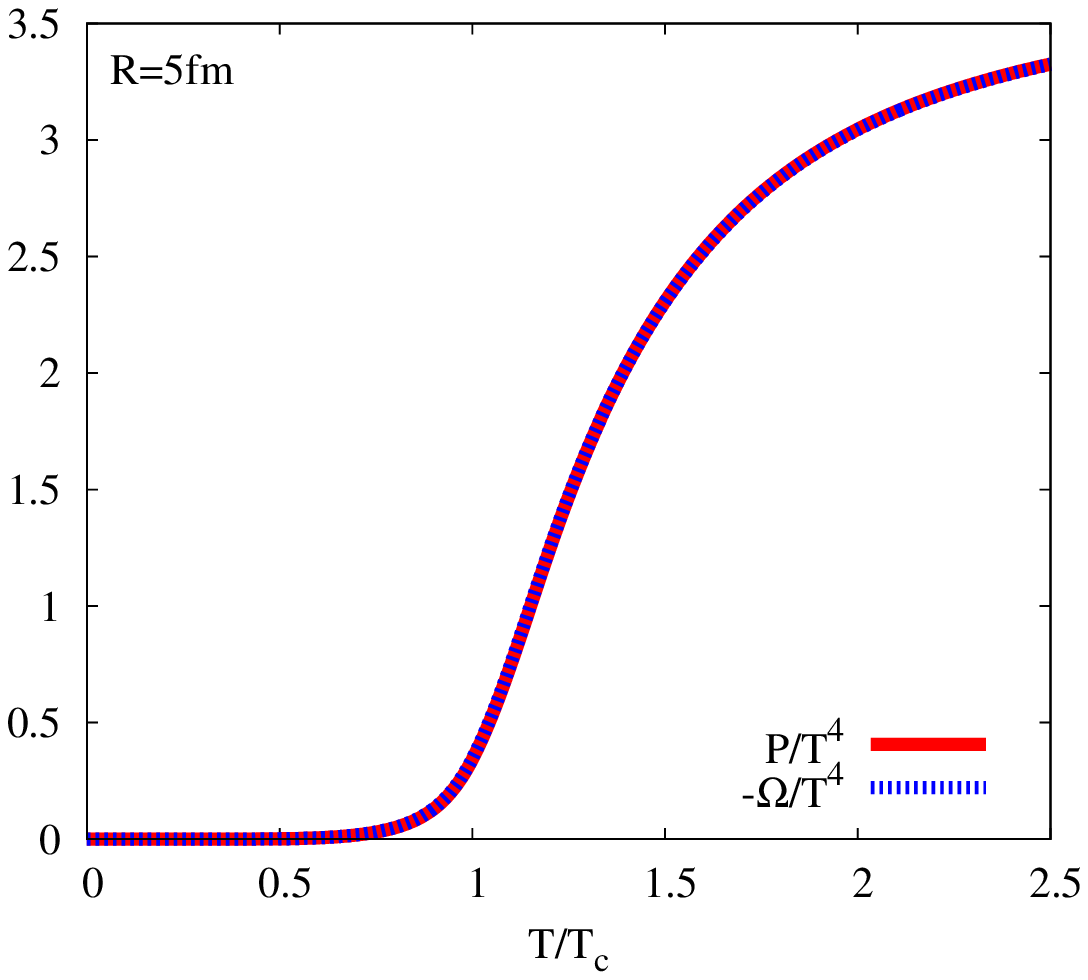}
\caption{(Color online) Pressure and energy density as
a function of temperature (top panel). Comparison of pressure and
$-\Omega$(bottom panel). These plots are for 2 flavor matter only.} 
\label {prs}
\end{figure}

The pressure inside a volume $V$ may be written as,
$P(T,\mu_q)=-\frac {\partial (\Omega (T,\mu_q)V) }{\partial V}$,
where $T$ is the temperature and $\mu_q$ is the quark chemical potential.
In the top left panel of Fig. \ref{prs} we plot the temperature 
dependence of scaled pressure ($P/T^4$) for 2 flavor system. As can be
seen there is a significant change in scaled pressure for small system
sizes. For example at $T_c$ the $P/T^4$ for a system with $R=2~fm$ is
almost half of that of an infinite system. As the temperature increases
the difference slowly diminishes. The decrease of scaled pressure with
increasing volume may be a surprise given that the constituent quark
masses were shown to decrease drastically with decreasing volume, which
should imply increase in pressure. This can be understood as follows.
With decreasing volume, not only the constituent masses decrease, but
also the lowest momentum increases due to the infrared cut-off. These
two conditions somehow seem to keep the lowest available energy of the
quark quasi-particles almost same for different volumes. Thus the
pressure does not increase with decreasing volume. However when plotted
against $T/T_c$ it seems to decrease because the $T_c$ itself is smaller
for smaller volumes, and therefore the pressure at the corresponding
$T/T_c$ for smaller volume is smaller than that for a larger volume.

The volume dependence is also quite strong for the energy density
$\epsilon={-T^2 \frac{\partial {(\Omega/T)}}{\partial T}}\Big|_V 
        ={-T \frac{\partial \Omega}{\partial T}}\Big|_V+ \Omega$.
In the top right panel of the Fig. \ref{prs}
we have plotted the $\epsilon/T^4$ as a function of $T/T_c$ for 
different system sizes. It has similar characteristics as $P/T^4$ but
the difference seems to diminish faster with increasing temperature.
As the system size becomes $R=5~fm$ both the scaled pressure and scaled
energy density converge to the $R \rightarrow \infty$ case for almost
all temperatures.

 It is well known that for infinite volumes the definition of pressure
simplifies to, $P(T,\mu_q)=-\Omega (T,\mu_q)$, which is commonly
used in the literature for PNJL models at infinite volumes. However
since we are considering finite volumes here it would be interesting
to check how much difference will it make if we keep using this
definition rather than the correct one with a volume derivative. In the
bottom two panels of Fig. \ref{prs} we have made a comparison of 
$-\Omega/T^4$ and $P/T^4$. For $R=2~fm$ we see that these two quantities
differ by about $10 \%$. Again, as the size goes close to $R=5~fm$ this
difference is almost washed out. 

\begin{figure}[htb]
\centering
\includegraphics[scale=0.5]{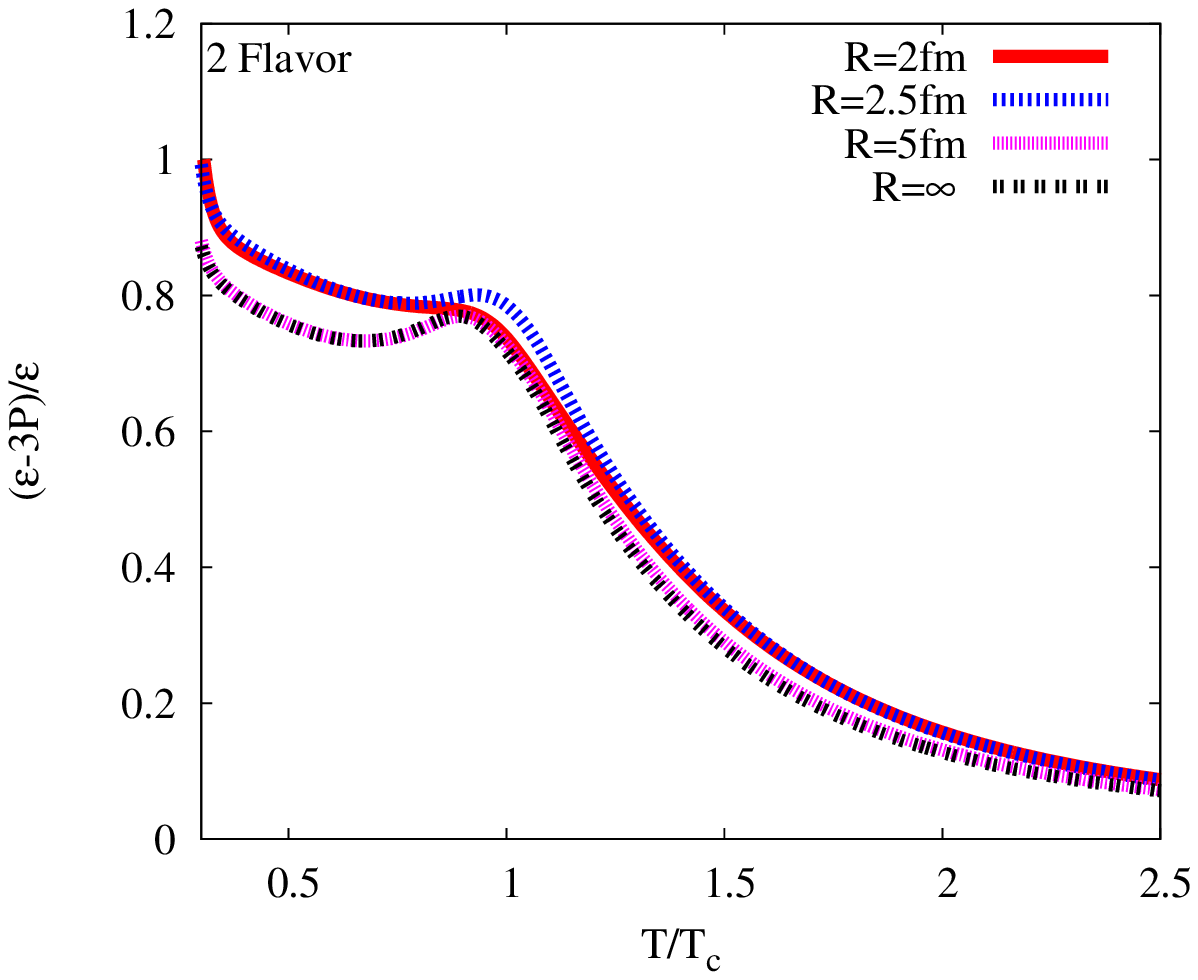}
\includegraphics[scale=0.5]{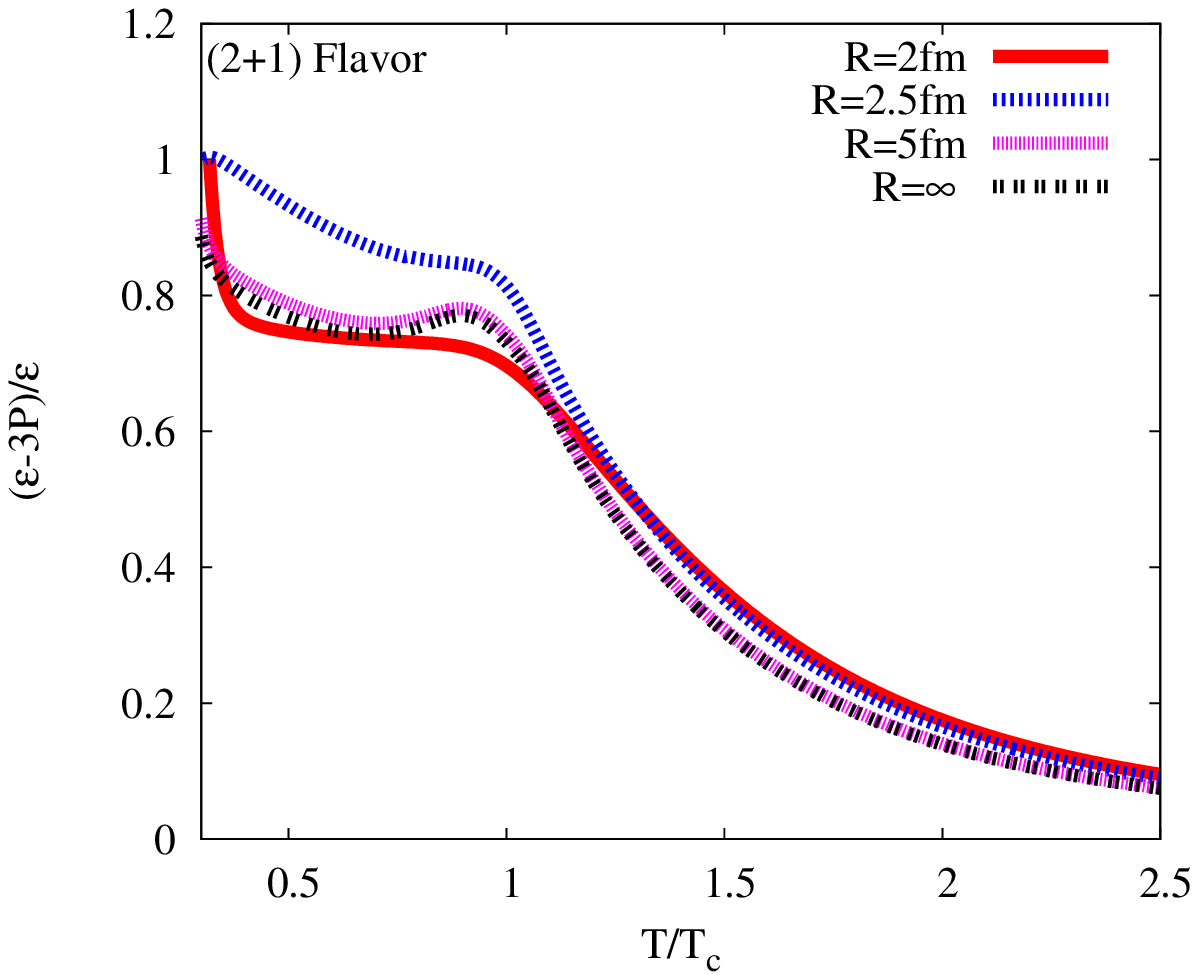}
\caption{(Color online) Conformal measure $\cC=1-3P/\epsilon$ for
different system sizes as a function of temperature.}
\label {e3p}
\end {figure}

Let us now consider the quantity $\epsilon-3P$. In our mean field
approach this is the trace of the energy-momentum tensor given by,
$\cT_{\mu \mu}= \epsilon-3P$. In a conformaly symmetric theory,
for example a theory of free massless quarks and gluons the 
energy momentum tensor is supposed to be zero as it signifies the
conservation of the conformal currents. Thus  $\epsilon = 3P$ in
that limit. In QCD however the conformal symmetry is broken due to
non-zero quark masses as well as quantum anomalies as evident from
the presence of a {\it scale} in the running coupling
constant~\cite{joglekar,neilsen}. Thus the energy-momentum tensor
does not remain traceless. This was also found to be true in the PNJL
model that have been reported in our earlier studies and compared with
LQCD results~\cite{ray1,ray2,deb2}. The PNJL model
is however not QCD and the reason for the scale symmetry breaking
is the introduction of an ultraviolet cut-off in the NJL part, a 
temperature scale $T_0$ in the Polyakov loop part and of course
a quark mass term similar to that in QCD. The physical implication
of the two different scales in the quark and Polyakov sector is to
give rise to separate crossover temperatures for the two sectors. 
To compare quantities obtained in PNJL model against LQCD
results one then averages out two crossover temperatures as done by
us here in the last section. Now for finite system sizes we have
introduced an infrared cutoff which should further enhance the effect
of conformal symmetry breaking. In fig. \ref{e3p} we show the
variation of the conformal measure $\cC={(\epsilon-3P)}/\epsilon$ with
temperature for both 2 flavor and and 2+1 flavor matter for different
system sizes. That the smaller system sizes lead to larger conformal
symmetry breaking effects is evident, except for the anomalous 
behavior of the lowest size of $R=2~fm$. (Though not shown in the
figure we found that the anomalous behavior starts at a size between
$2~fm \le R \le 2.5~fm$. The reason for this behavior is not clear at the
moment and requires further investigation.)

\begin{figure}[htb]
\centering
\includegraphics[scale=0.5]{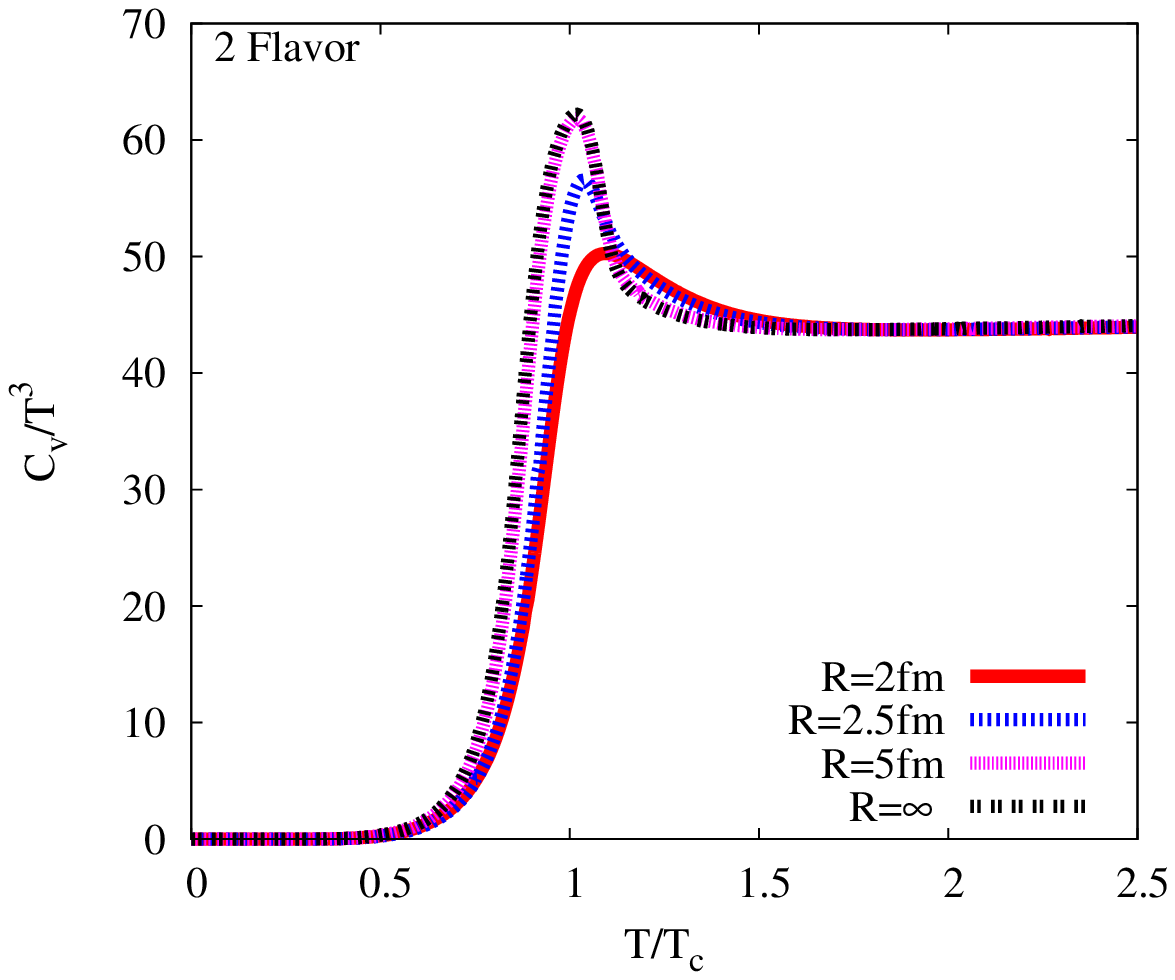}
\includegraphics[scale=0.5]{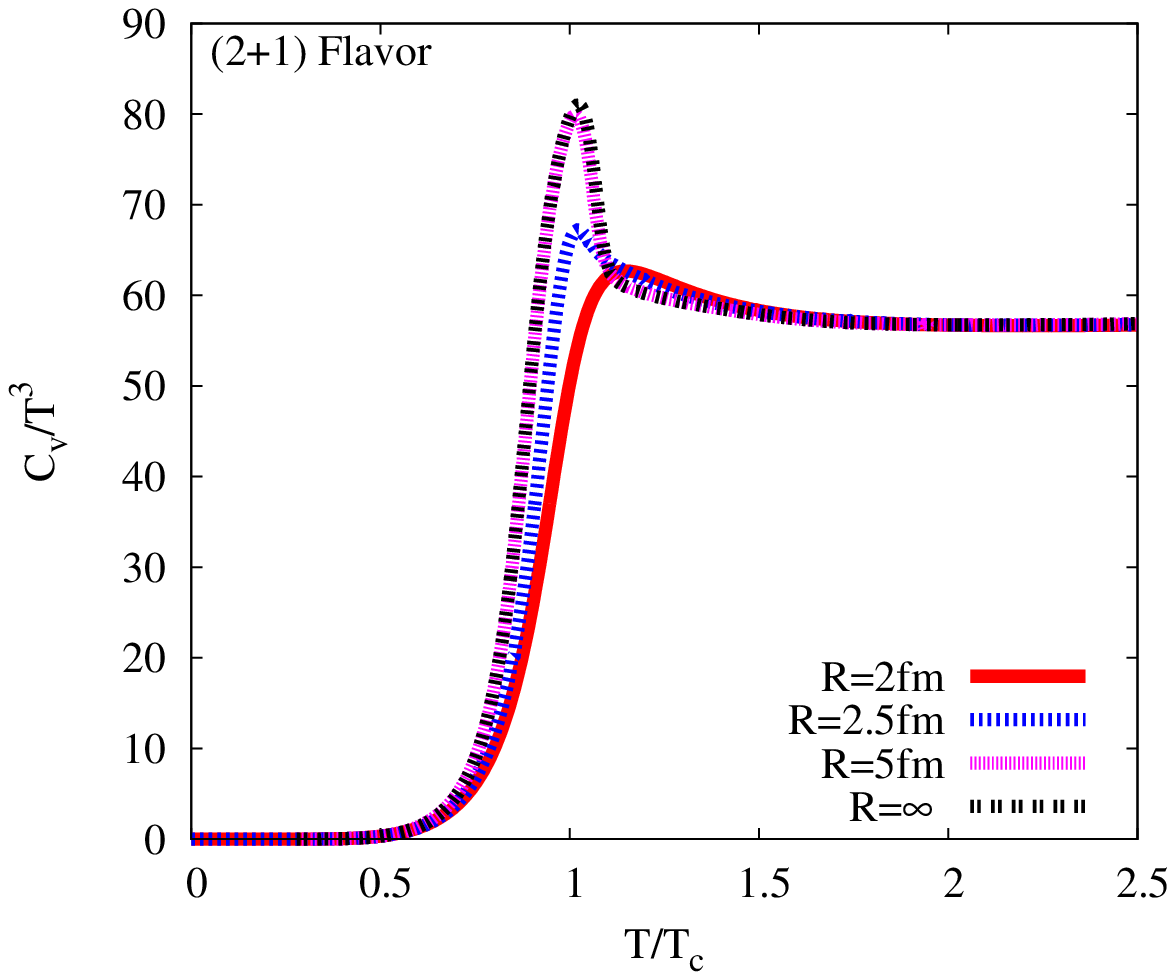}
\caption{(Color online) 
Variation of specific heat with temperature for different system sizes.}
\label {cv}
\end{figure}

The specific heat at constant volume 
$C_V=\frac{\partial \epsilon }{ \partial T} \Big|_V$ 
is shown in Fig. \ref{cv}. We find that with the change in volume,
$C_V$ changes prominently up to the temperature corresponding to the
crossover region. For smaller volumes the specific heat is smaller
indicating a higher rise in temperature for the same rise in energy
density. Obviously this can be correlated with the temperature
dependence of energy density discussed earlier in Fig. \ref{prs}.
We found that a given amount of scaled energy density is obtained
at a higher scaled temperature for a smaller volume. This can be of
interest in heavy-ion collision experiments. A given energy density
deposited in the finite volume would create a plasma with temperature
somewhat higher than that expected in a similar volume inside an
infinite volume system having the same energy density.

The specific heat is also a measure of energy fluctuations in the
system~\cite{korus}. Fluctuations tend to rise sharply near a phase
transition. For a crossover they are somewhat subdued. Obviously
for finite volumes a true phase transition is not possible and
as one keeps on decreasing the volume all signatures even for a
crossover should die down. This is exactly the behavior of $C_V$
as presented in Fig. \ref{cv}.

\begin{figure}[htb]
\centering
\includegraphics[scale=0.5]{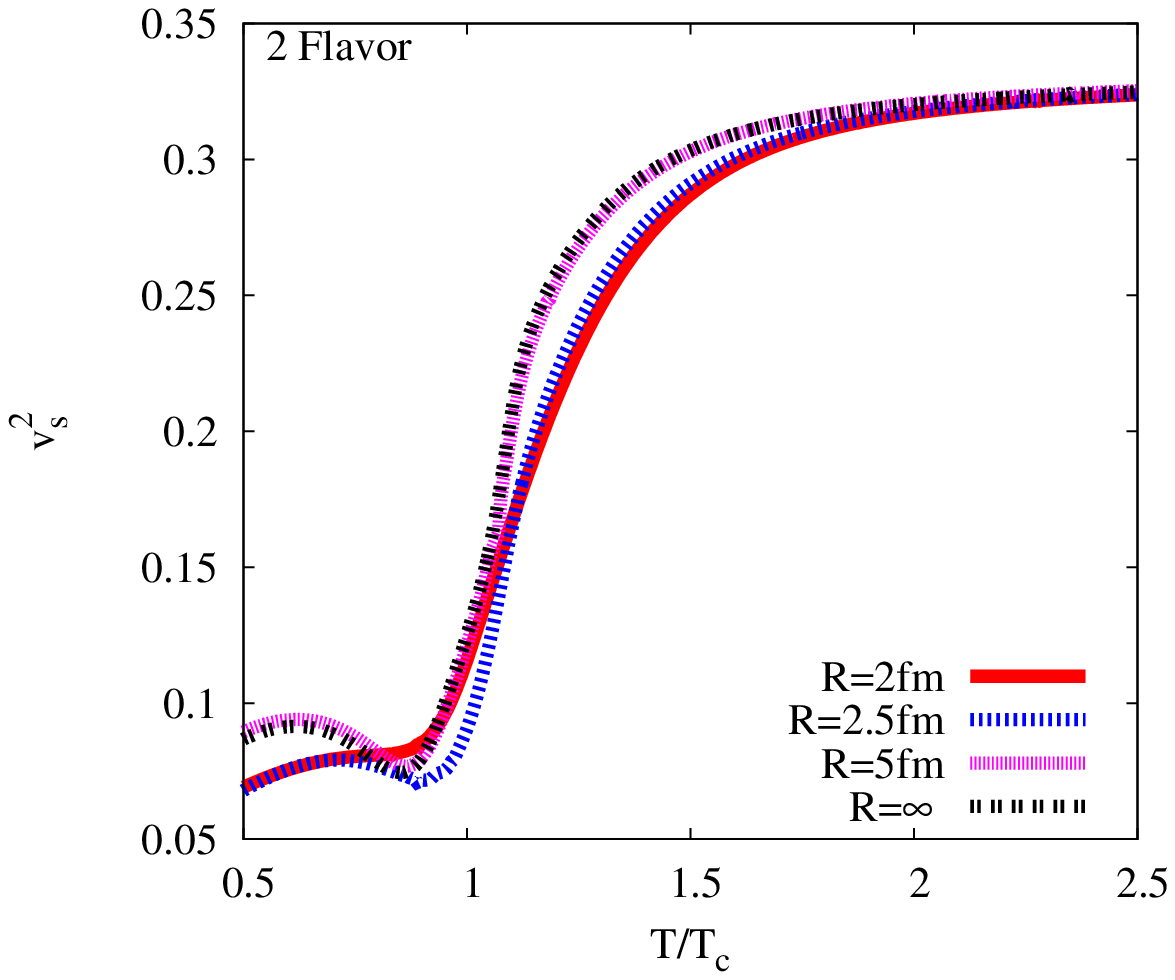}
\includegraphics[scale=0.5]{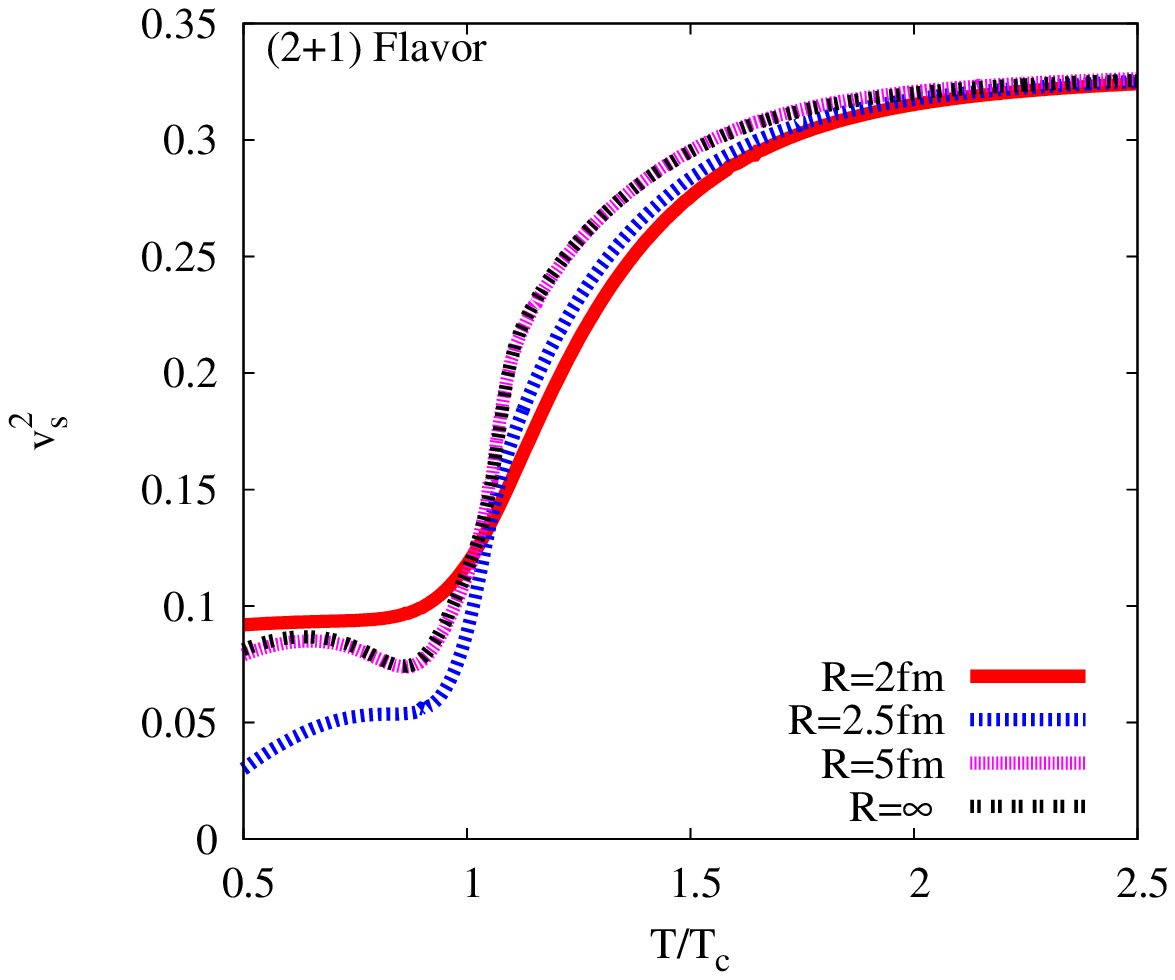}
\caption{(Color online) 
Variation of squared speed of sound with temperature for different
system sizes.}
\label {cs}
\end{figure}

The squared speed of sound
$v_s^2 =  \frac{\partial P }{ \partial \epsilon}$ is shown in 
Fig. \ref{cs}. At large temperatures the $v_s^2$ reaches its
maximum value as the system becomes almost ideal. Interactions
grow with decreasing temperatures resulting in the lowering of
$v_s^2$. The conformal measure $\cC$ may be considered as
a measure of the strength of the interaction in the system.
Thus lower the value of $\cC$, higher should be the value of
$v_s^2$. This is evident from Fig. \ref{cs}, where we find
the $v_s^2$ to decrease with decreasing temperature, just opposite
to the behavior of $\cC$ shown in Fig. \ref{e3p}. This correlation
between $\cC$ and $v_s^2$ also apparent for variation in volume.
With decreasing volume the speed of sound decreases. (In fact an
anomalous behavior for the smallest size $R=2~fm$ is also apparent
for $v_s^2$.) A smaller speed of sound for smaller volumes would
mean a slower flow for finite size systems created in heavy-ion
collisions.

\vskip 0.3in
{\section {Properties of Non-strange mesons} 
\label{meson}}

For infinite volumes the meson properties in the PNJL model has been
discussed for 2 flavors~\cite{hansen,fuliu} as well as for 2+1
flavors~\cite{deb,costa}. In this section we describe the
properties of non-strange mesons at finite volumes in the PNJL model.
A detailed account of the calculational procedure for meson masses
at finite temperatures and densities in the PNJL model may be found
in Ref.~\cite{deb}. Here we sketch the outline of the task. 
 
The collective excitations, the fluctuation of the mean field around
the vacuum can be handled within the Random Phase Approximation
(RPA)~\cite{fett}.  In this approximation, which is equivalent to
summing over the ring diagrams, the retarded correlation function 
for a meson $M$ is given by, 
\begin {equation}
 D_M^R=\frac{\Pi^M}{1-2G_M{\Pi^M}}.
\label{eq.rpa}
\end {equation}
Here $G_M$ is the suitable coupling constant and $\Pi_M(k^2)$ is the
one-loop polarization function for the mesonic channel under
consideration. Within the RPA, $\Pi^M$ may be written as,
\begin{equation}
  \Pi^{M} \equiv \idp \Tr \left[\Gamma_M S(p+q) \Gamma_{M} S(q)\right], 
\label{eq.pim}
\end{equation}
where $S(p)$ is the Hartree quark propagator, $\Gamma_M$ is the
appropriate combination of gamma matrices of different mesonic channels
and the trace is taken over the Dirac and color spaces. The lower 
limit on the integration is now required for finite volume studies. 

Here we concentrate on the scalar $(\sigma)$ and pseudoscalar
$(\pi)$ channels. These contributions can be written as,
\begin{eqnarray}
 \Pi^{ab}_\pi (q^2) = \idp \Tr
   \left( i \gamma_5 \tau^a S(p+q) i \gamma_5 \tau^b  S(q) \right)
\nonumber\\
 \Pi_{\sigma} (q^2) = \idp \Tr \left( S(p+q) S(q) \right).
\end{eqnarray}
The pole mass can be obtained by solving,
\begin{equation}
   1-2G_M{\Pi^M}(\omega = m_M, {\vec q}=0)=0.
\label{eq.pole}
\end{equation}
where $m_M$ is the mass of a particular meson. The detailed expression
for $\Pi^M$ and $G_M$ for $\pi$ and $\sigma$ mesons may be found in 
Ref.~\cite{deb}.

\begin{figure}[htb]
\centering
\includegraphics[scale=0.5]{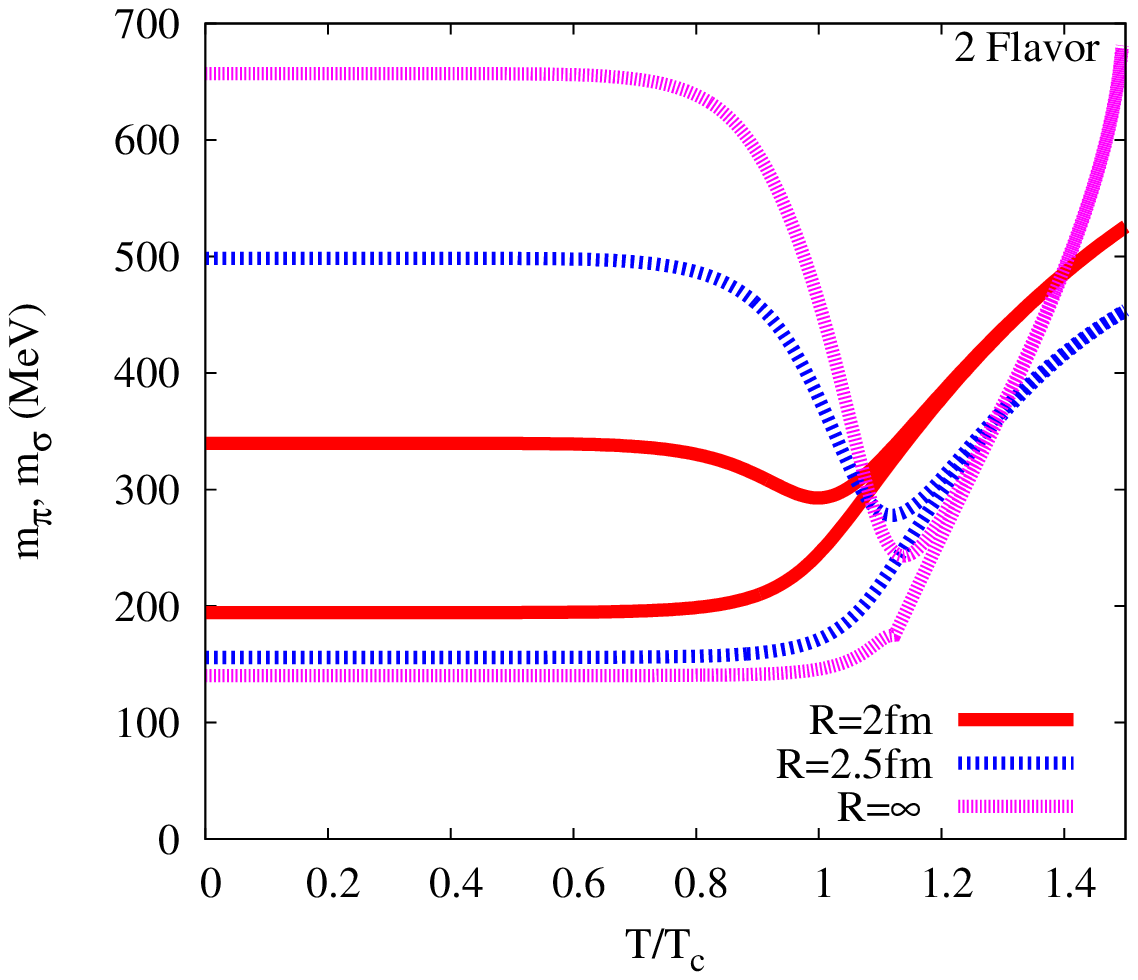}
\includegraphics[scale=0.5]{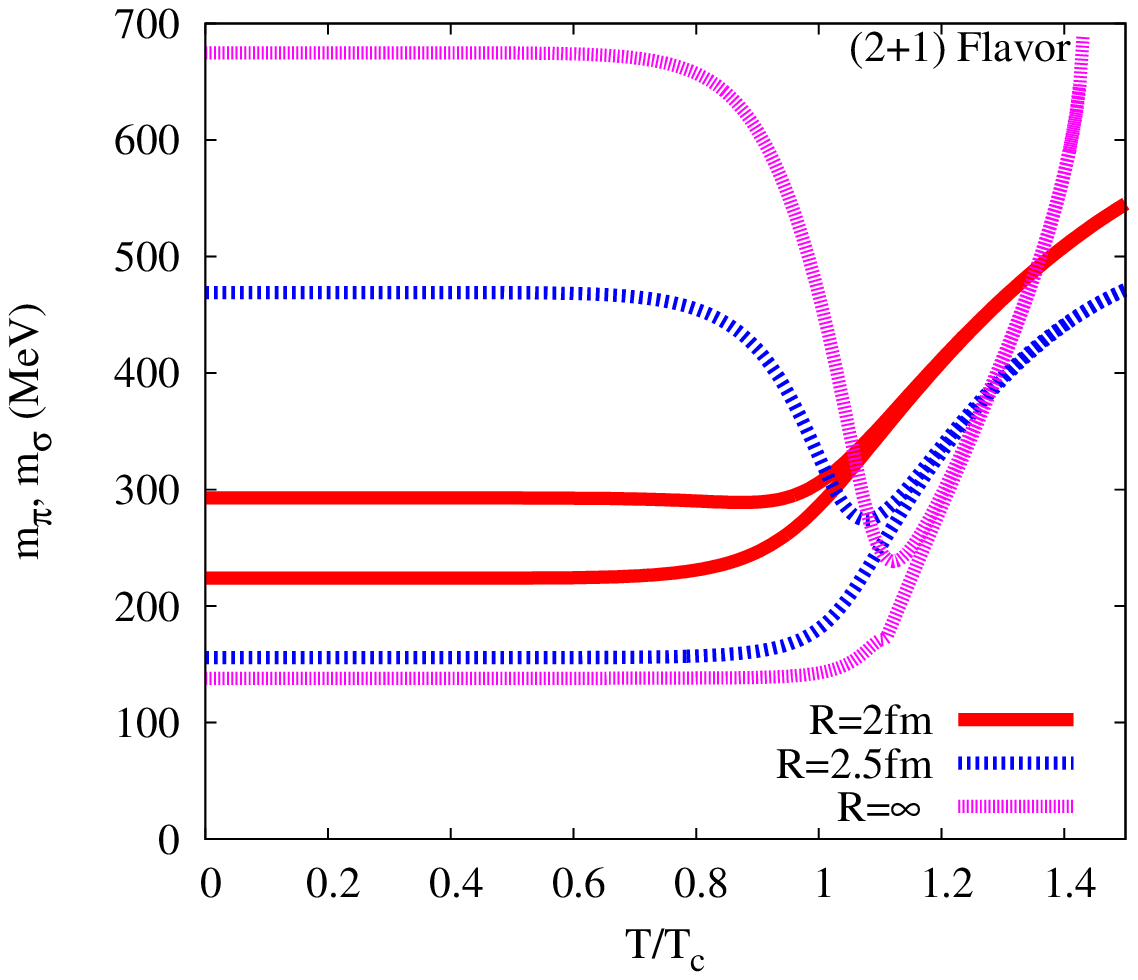}
\includegraphics[scale=0.5]{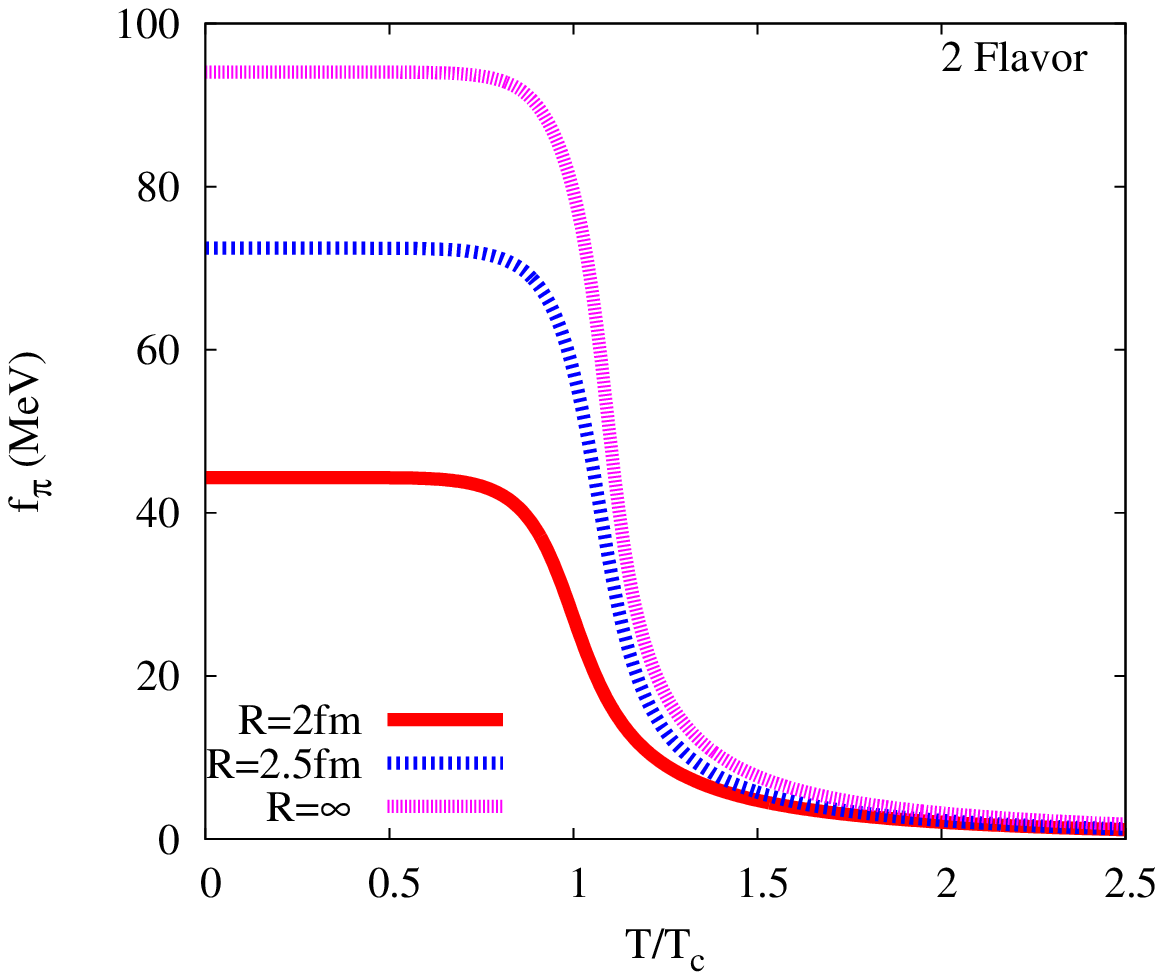}
\includegraphics[scale=0.5]{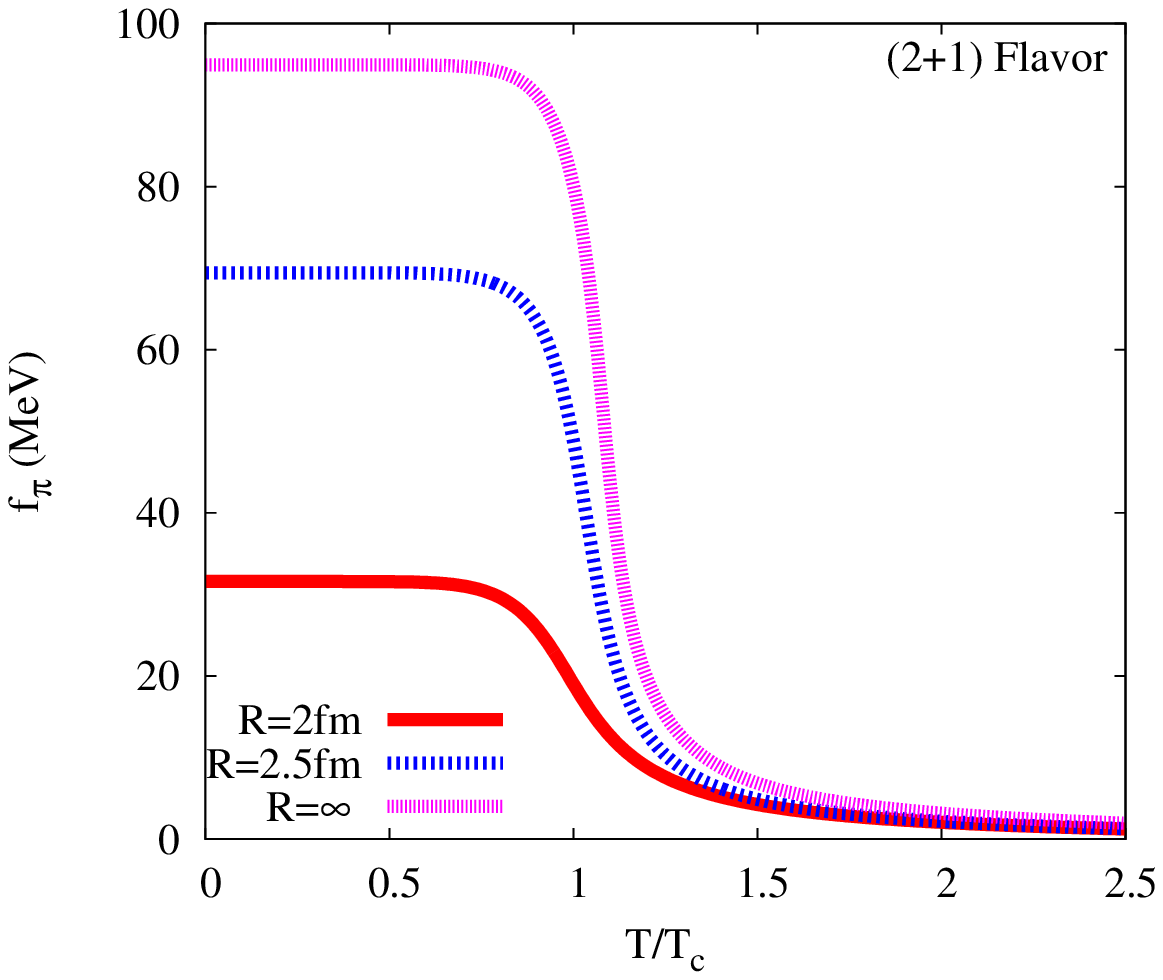}
\caption{(Color online) 
Variation of meson masses (upper panel) and pion decay constant
(lower panel) with temperature for different system sizes. In the
upper panel the three monotonously rising curves are for $m_\pi$ and
the other three are for $m_\sigma$.}
\label {mesonmass}
\end{figure}

In the upper panels of Fig. \ref{mesonmass} we have plotted the masses
of pion ($m_{\pi}$) and sigma ($m_{\sigma}$) as a
function of temperature for different system sizes. In any given volume
we see that for low temperatures the masses of pion and sigma are
different and they become degenerate above $T_c$ where chiral symmetry is
expected to get restored. With decrease in volume we find the pion mass
to increase. However above $1.2~T_c$ the pion mass for infinite volume
suddenly shoots up above those for the finite volumes. This may have
important consequences in heavy-ion reactions where system size is small.
Whereas for infinite volume the fast increasing mass of pion would
drastically reduce chances of obtaining pion-like bound states, the same
may not be true for finite volume. We note here that the increase of
pion mass with decreasing volume has also been observed in computations
with chiral perturbation theory \cite{chipt} and renormalization group
methods in quark-meson model \cite{renqm}.

While the mass of pion increases with decreasing volume at low
temperatures, the mass of sigma is found to decrease 
quite fast. One can actually see a trend to the masses of the two 
chiral partners becoming closer to each other with decreasing volume.
This, yet again, shows that chiral symmetry breaking effects reduce with
decreasing volumes.

The pion decay constant may be obtained from the matrix element
$\langle 0 | J_{\mu,5}^a |\pi^b(k) \rangle =
                                          i \delta_{ab} f_\pi k_\mu$,
where
$J_{\mu,5}^a={\bar \psi} \gamma_\mu \gamma_5 \frac{\tau^a}{2} \psi$ is
the chiral current. At finite temperature and for a particular volume
it can be written as (see~\cite{klev,hat1} and including the low
momentum cut-off $\lambda$),
\begin{equation}
f_\pi^2 = \frac{3 M_u^2}{2\pi^2}
 \left[ \int_\lambda^\Lambda \frac {p^2dp}{E_p^3} -
       2 \int_\lambda^\infty \frac {p^2dp}{E_p^3}  f(E_p)\right]
\end{equation}
where $E_p = \sqrt {p^2 + M_u^2}$ is the single particle energy of a
light quark and $f(E_p)$ is the distribution function properly modified
due to Polyakov loop interaction. 

As shown in the lower panels of Fig. \ref{mesonmass}, the pion decay
{\it constant} decreases both with the decrease in temperature and with
that of system size. The decrease of $f_\pi$ with temperature has also
been observed in other effective models~\cite{fuliu,fpi_model},
Dyson-Schwinger approaches~\cite{fpi_dys} as well as in LQCD~\cite{fpi_lat}. 
This is also an indication of the restoration of chiral
symmetry as $f_\pi$ is directly proportional to the divergence of the
chiral current.

The tendency of chiral symmetry getting restored in finite volumes may
also be noted by comparing
Fig. \ref{mesonmass} with Fig. \ref{fig_mass}. At low temperatures the
constituent quark masses decrease with decreasing volume. It so happens
that the light constituent quark masses become smaller than the pion
mass for the smallest sizes studied here. These quarks should then
become thermodynamically more favored than the pions. Though fortunately
in the PNJL model, such constituent quarks will be suppressed due to
the presence of the Polyakov loop, the pions would still loose their
significance as the lightest particles that made them suitable
candidates for becoming the Goldstone bosons. Thus what seems to 
happen is that the decrease of volume restores the spontaneous 
breaking of chiral symmetry in the same way as increase in temperature.
The {\it critical size} $R_c$ for such symmetry restoration would be
somewhere between $2~fm$ and $2.5~fm$. From Fig. \ref{fig_mass} it
may be noted that this range of sizes is almost equal to the 
respective constituent quark masses. This observation is commensurate
with the expectation from chiral perturbation theory that chiral
symmetry restoration may take place once the quark masses become equal
to the inverse of system size~\cite{gasser}.

\begin{figure}[htb]
\centering
\includegraphics[scale=0.5]{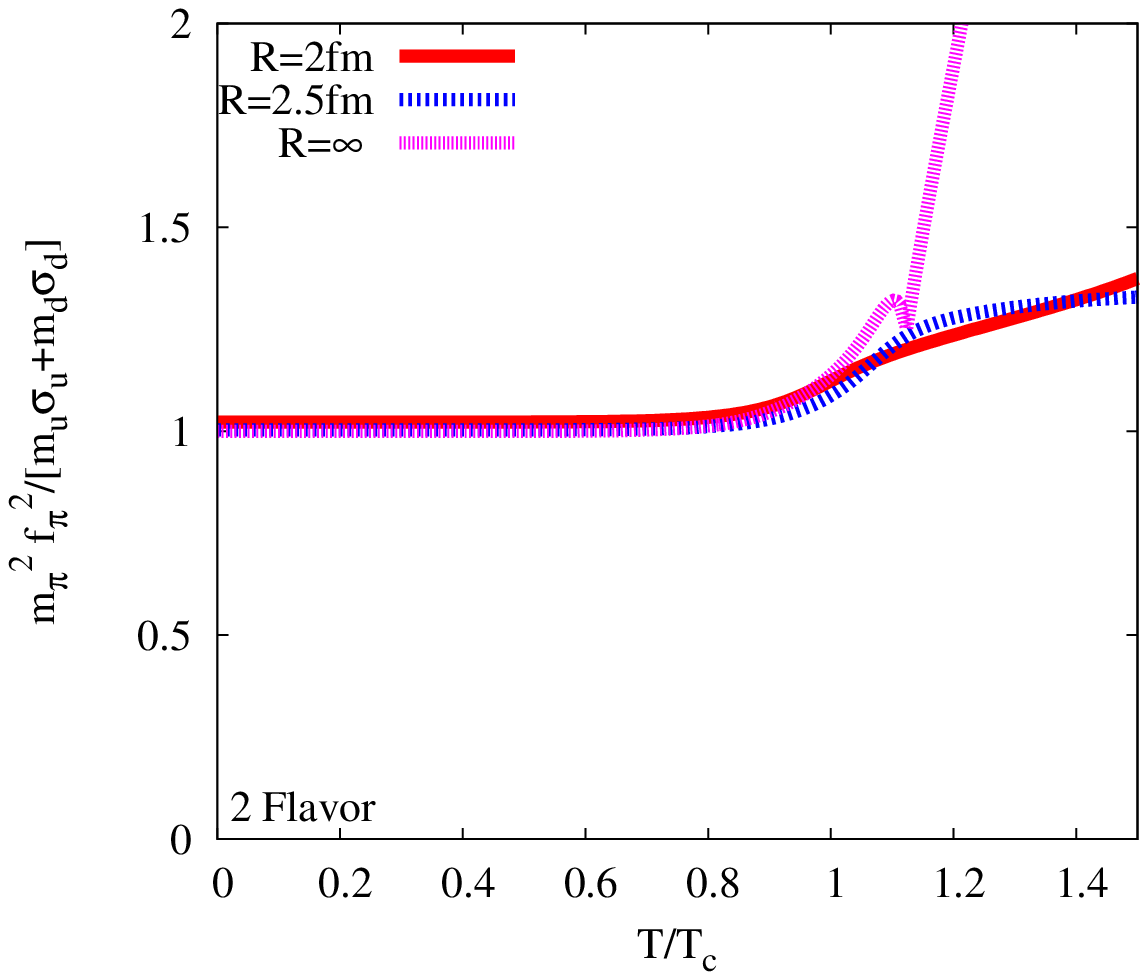}
\includegraphics[scale=0.5]{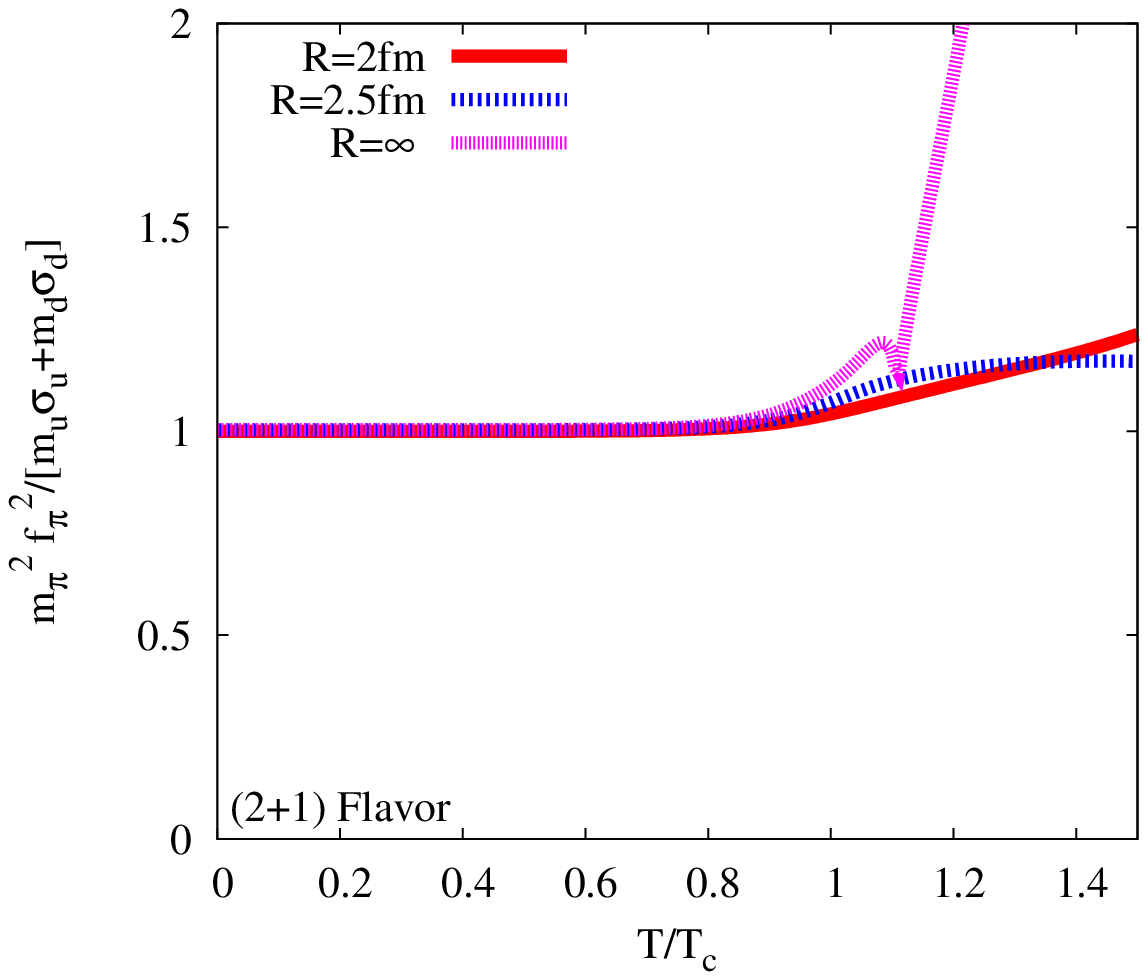}
\caption{(Color online) Checking for violation of the GMOR relation
as a function of temperature and system size.}
\label {fig.gmor}
\end {figure}

With all the strong indication of a possible chiral symmetry
restoration with decreasing volume it would be interesting to see what
happens to the Gell-Mann Oakes Renner (GMOR)~\cite{gmor} relation,
which in the lowest order of chiral expansion is given by
$f_\pi^2 m_\pi^2=m_u \sigma_u+m_d \sigma_d$.
Normally with increase in temperature as the spontaneously broken
part of the chiral symmetry gets restored the GMOR relation should
start to break down. This is exactly what we find in our calculations
and shown in Fig. \ref{fig.gmor}. But surprisingly we find that
similar effect is not observed for the decrease in volume and the GMOR
relation holds good for the all the ranges of volumes we considered
almost up to temperatures as high as $0.8~T_c$. In fact even at higher
temperatures the GMOR relation is violated the most for infinite
volumes. The way one can understand this is that for a physical
chiral expansion a quantity $m_\pi/\cM$ is required, where $\cM$
is some suitable scale. In chiral perturbation theory $\cM$ is usually
the neucleon mass, in zero temperature NJL or PNJL models it is the
high momentum cut-off $\Lambda$, etc. For finite temperatures one can
then consider $T$ to play the role of $\cM$. Thus, given a temperature
if the corresponding $m_\pi$ in a given volume is less than $T$,
chiral identities would work properly (see e.g.~\cite{son}). So here
we have a situation where chiral symmetry is getting restored while
partial conservation of axial current is still maintained.

\vskip 0.3in
{\section {Conclusion}
\label{secconclu}}

We have tried to understand the dynamics of strongly interacting matter
inside finite volume in the framework of PNJL model with saddle point
approximation. Several interesting results were observed that can have
important implications for heavy-ion collision experiments. Our major
finding was that the spontaneously broken chiral symmetry may be
restored at much lower temperatures in small volume. This was shown
through the computation of various thermodynamic observables as well
as certain hadron properties.

Changes in the equation of state and speed of sound may have important
consequences in the flow properties of the exotic medium created in the
experiments. A measure of the specific heat in heavy-ion experiments is
the transverse momentum fluctuations. We find the specific heat to
decrease with decreasing volume indicating that the momentum
fluctuations may not be as large as expected at a given $T/T_c$.

From the variation of the phase boundary with changing volume we
demonstrated a stronger possibility of finding the signatures of
a critical end point in low energy experiments that intend to create
high baryonic densities where the expected temperature is not too
high.

Finally from the hadron properties we observed the possibility
of obtaining a chiral symmetric but confined phase in small
volumes. We hope that a combination of heavy ion collisions and
not-so-heavy ion collisions at similar center of mass energies,
followed by an appropriate finite size scaling study may give us
a better understanding of the QCD phase structure.

As discussed earlier we made a couple of simplified assumptions
in this work. The Polyakov loop potential used here does not have an
explicit volume dependence. The discrete momentum states in the quark
potential was replaced with a continuum, and the only explicit
dependence on system size was through the lower momentum cut-off.
Though we believe that these assumptions would not affect the gross
features observed, we hope to address these issues in future. It would be highly 
desirable to have a concurrent study of finite size effects in Polyakov-Quark-Meson 
models \cite{pqmsch} to further understand the systematics of model artifacts. 

\vskip 0.3in
\acknowledgments{
We would like to thank Sanatan Digal, Tamal K. Mukherjee and Ajit M.
Srivastava for many useful discussions and comments.
A.B. thanks  UGC  (UPE and DRS) and DST and R.R. thanks DST for support.
}

\end{document}